\documentclass[draftcls,onecolumn]{IEEEtran} \def\onetwocol#1#2{#1}

\usepackage{cite}      

\usepackage{subfigure}
\usepackage{amsmath,mathrsfs} 
\usepackage{amssymb} 
\usepackage{amsfonts}
\usepackage{mathtools}

\usepackage{epstopdf}
\usepackage{ifthen}

\usepackage{amsmath}   
\newcommand{\inZ}{\in{\mathbb Z}}
\newcommand{\inR}{\in{\mathbb R}}

\newcommand{\C}{ \mathbb{C}}
\newcommand{\R}{ \mathbb{R}}

\newcommand{\Lop}{{\rm L}}
\newcommand{\Pop}{{\rm P}}
\newcommand{\Top}{{\rm T}}
\newcommand{\Dop}{{\rm D}}
\newcommand{\Iop}{{\rm I}}
\newcommand{\Identity}{{\rm Id}}
\newcommand{\dint}{{\rm d}}

\newcommand{\Fourier}{ \mathcal{F}}

\newcommand{\bw}{{\boldsymbol \omega}}
\newcommand{\s}{{\zeta}}

\newtheorem{definition}{Definition}
\newtheorem{proposition}{Proposition}
\newtheorem{corollary}{Corollary}
\newtheorem{property}{Property}

\newtheorem{theorem}{Theorem}

\def\Spc#1{{\mathcal{#1}}}  

\def\One{\mathbf{1}}    

\def\Form{ \widehat {\mathscr{P}}}        
\def\E{\mathbb E}        
\def\Corr{\mathcal B}        

\def\dint{\;\mathrm{d}}

\begin{document}

\title{A unified formulation of Gaussian vs. \onetwocol{\\}{\\} sparse stochastic processes---\\
Part II: Discrete-domain theory}

\author{Michael~Unser,~\IEEEmembership{Fellow,~IEEE,} Pouya Tafti,~\IEEEmembership{Member,~IEEE,} Arash Amini, and Hagai Kirshner. 
\thanks{The authors are with the Biomedical Imaging Group (BIG), \'Ecole Polytechnique F\'ed\'erale de Lausanne (EPFL), CH-1015 Lausanne, Switzerland. }
}

\onetwocol{ }{\markboth{IEEE Transactions on Signal Processing,~Vol.~X, No.~XX,~2012}{
}}
%



\maketitle

\begin{abstract}
This paper is devoted to the characterization of an extended family of CARMA (continuous-time autoregressive moving average) processes that are solutions of stochastic differential equations driven by white L\'evy innovations. These are completely specified by: (1) a set of poles and zeros that fixes their correlation structure,  and (2) a canonical infinitely-divisible probability distribution that controls their degree of sparsity (with the Gaussian model corresponding to the least sparse scenario).
The generalized CARMA processes are either stationary or non-stationary, depending on the location of the poles in the complex plane.
The most basic non-stationary representatives (with a single pole at the origin) are the L\'evy processes, which are the non-Gaussian counterparts of Brownian motion. We focus on the general analog-to-discrete conversion problem and introduce a novel spline-based formalism that greatly simplifies the derivation of the correlation properties and joint probability distributions of the discrete versions of these processes. We also rely on the concept of generalized increment process, which suppresses all long range dependencies, to specify an equivalent discrete-domain innovation model. A crucial ingredient is the existence of a minimally-supported function associated with the whitening operator $\Lop$; this B-spline, which is fundamental to our formulation, appears in most of our formulas, both at the level of the correlation and the characteristic function. We make use of these discrete-domain results to numerically generate illustrative examples of sparse signals that are consistent with the continuous-domain model.

\end{abstract}

\section{Introduction}
In our companion paper, we have set the foundations of a general innovation framework that leads to the specification of a broad class of continuous-time stochastic processes\cite{Unser2012}. 
The powerful aspect of the formulation is that it unifies the classical theories of stationary Gaussian processes \cite{Papoulis1991}, on the one hand, and L\'evy processes on the other \cite{Sato1994}, the idea being that
these processes can all be generated by applying a proper integral operator ($\Lop^{-1}$) to some admissible (white) innovation process. We have also shown that switching to a non-Gaussian excitation (within the class of admissible solutions) necessarily induces a sparse behavior. An intriguing consequence of the latter is that it improves the performance of wavelet-like transformations: in the non-Gaussian regime, these tend to provide better $N$-term signal approximations than the classical KLT (or the DCT) does, 
which is the reverse of what happens in the classical Gaussian setup (cf. \cite[Sections II, V.D]{Unser2012}).
This suggests that this type of modeling is highly relevant for modern signal processing, which is presently very much focused on the design of signal recovery algorithms that promote sparsity in some transformed domain.
While the proposed generation mechanism is remarkably simple conceptually, it is not quite as straightforward to formulate rigorously because the underlying innovations (admissible white noise excitations = L\'evy noise) can only be properly defined in the sense of distributions \cite{Gelfand-Villenkin1964,Yaglom1986}.
Statisticians usually work around the difficulty by defining processes through stochastic integrals (It\^o calculus) which avoids the explicit reference to white noise \cite{Ito:1951,Okensal2007}; the downside of this widely-used framework is that it partly hides the system-theoretic aspects.
\begin{figure}
\centering
\includegraphics[width=8cm]{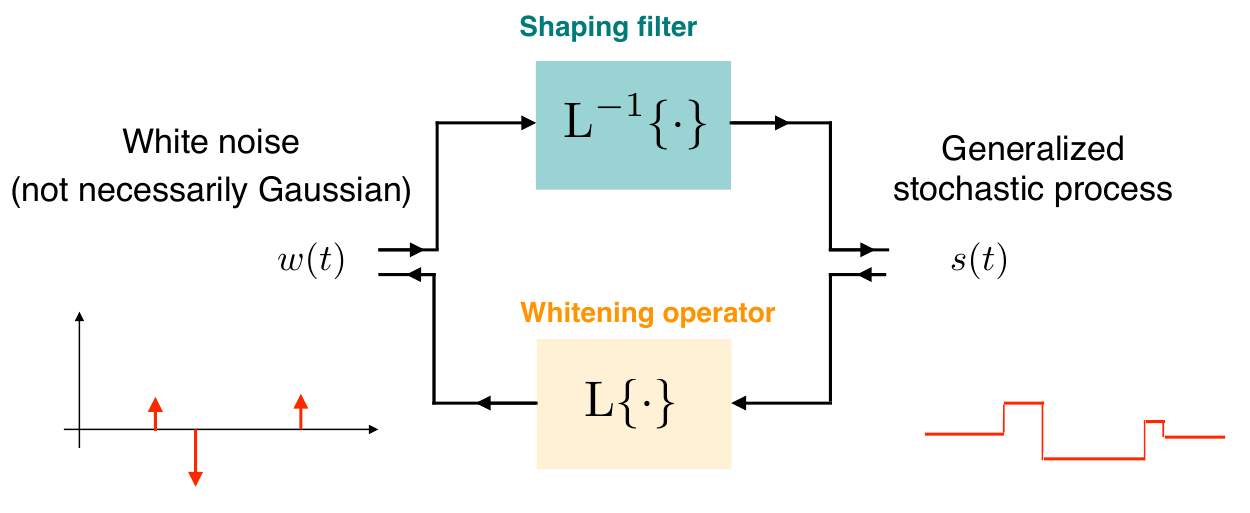}
  \caption{\label{fig:innovation}
Innovation model of a generalized stochastic process. The process is generated by application of the inverse operator $\Lop^{-1}$ to a continuous-domain white noise process $w$. The generation mechanism is general in the sense that it extends to the complete family of (non-Gaussian) noises $w=\dot W$ that formally correspond to the weak derivative of some classical L\'evy process $W(t)$. Gaussian processes are recovered by taking $W(t)$ to be the Wiener process (a.k.a. Brownian motion). The output process $s(t)$ is stationary iff. $\Lop^{-1}$ is shift-invariant.}
\end{figure}

The innovation model described in Fig. 1 is attractive to engineers because it establishes a direct link between stochastic processes and linear system theory. It also suggests that it is possible to transpose some standard deterministic techniques (e.g., determination of impulse responses, filtering, sampling of signals, cardinal spline interpolation) to the stochastic setting, which is mostly what this work is about. In other words, once one has gone through the effort of properly defining and understanding the notion of a continuous-domain white L\'evy noise, the remaining characterization problem can be addressed by relying on the powerful (deterministic) tools of functional and harmonic analysis.
The non-trivial aspect is that one needs to resolve some instabilities (in the form of singular integrals), both at the system level to allow for non-stationary processes, and at the stochastic level because the most interesting sparsity patterns are associated with unbounded L\'evy measures (cf. \cite[Section III.D]{Unser2012}).

\begin{table}
\caption{Typology of continuous-time stochastic processes}
\centering
\begin{tabular}{c | l l}
 \hline  \hline \\[-1ex]
 & Gaussian & Sparse\\[2ex]
\hline\hline\\[-2.5ex]
{Stationary} & {classical ARMA theory} & Non-Gaussian \\
& & CARMA processes \\[1ex]
\hline{Non-stationary} & Brownian motion & L\'evy processes \\
&  and present extensions &   and present extensions \\
 \hline
\end{tabular}
\end{table}

In the present paper, we investigate the discrete-time implications of the theory for the extended class of continuous-time processes which are ruled 
by ordinary differential equations (cf. the typology of processes shown in Table 1). The stationary Gaussian members of the family are well studied and
play a central role in traditional system modeling, signal processing and control theory \cite{Astrom1970,Papoulis1991,Gray2004}. There is also a well-known discrete connection in the sense that the sampled version of a Gaussian ARMA process is itself a discrete ARMA process with the discrete and continuous-domain poles being related by the exponential map: $\{z_n=e^{\alpha_n}\}_{n=1}^N$ \cite{Doob1962,Astrom1970,Wahlberg1993}.  Less obvious is the determination of the MA component of the discrete model which is jointly dependent upon the continuous-domain poles and zeros\cite{Kirshner2011}. Another classical instance is provided by the L\'evy processes, including Brownian motion,  which are commonly used in financial mathematics\cite{Black1973,Schoutens2003}. L\'evy process are especially interesting in that context because of their ability to replicate jumps in price assets \cite{Cont2004,Schoutens2003}. They are not as popular in signal processing circles, probably due to the fact that they are non-stationary; yet, it has been pointed out recently that they are actually very relevant because they are the processes for which some of present sparsity-based algorithms (e.g., TV-denoising) are statistically optimal\cite{Unser2011}. The final important subclass is made up of the so-called CARMA processes---the non-Gaussian extension of the classical ARMA processes \cite{Brockwell2001}. 
Special instances of such stationary processes have been applied to financial modeling\cite{Barndorff2001} and, to a lesser extent, signal processing \cite{Shao1993,Godsill2006,Yang2007}.

	In the sequel, we present a systematic characterization of the sampled versions of these processes. The primary contributions along the way are :
	\begin{itemize}
\item An addition to the non-stationary branch of the CARMA family via the introduction of generalized boundary conditions and ``regularized" inverse operators for the solution of unstable stochastic differential equations (SDE).
\item The specification of the {\em generalized increment process} which is a stationarized and ``localized'' version of the signal with the shortest possible range of dependencies.
\item The uncovering of the fundamental role of the exponential B-splines in the statistical characterization of the CARMA processes. Not only do such B-splines correspond to the autocorrelation function of the generalized increment processes, but they do allow for a remarkably concise description of the joint characteristic functions of the discrete versions of these processes.
\item The derivation of the discrete counterpart (finite difference equation) of the continuous-domain innovation model. The proposed formulation also extends to the non-Gaussian and/or non-stationary variants of these processes.

\end{itemize}

The paper is organized as follows. In Section II, we briefly review the general innovation model which specifies the broadest possible class of continuous-time linear stochastic processes. We also recall the inverse-operator method of solution which results in a complete characterization of the generalized CARMA processes\cite{Unser2012}. In Section III, we show how we can use finite-difference operators to partly decouple CARMA and generalized L\'evy processes.
In Section IV, we investigate the discrete-domain aspects of the theory by considering the sampled versions of these processes. In particular, we establish exponential spline-based interpolation formulas that connect the discrete and continuous-domain correlations of the CARMA processes.
We explicitly determine the $K$th-order characteristic function of the samples of the corresponding generalized increment processes, which are maximally decoupled. This naturally leads  to the specification of some equivalent discrete-domain ARMA-type innovation model. In Section V, we use those results in conjunction with exponential spline calculus to develop numerical algorithms for the generation of CARMA processes with a special attention to the non-Gaussian, non-stationary scenarios.  We conclude the paper with the presentation of illustrative examples of sparse processes in Section VI.

\section{Review of continuous-time results}
We start with a brief review and discussion of the key results of our theory of generalized stochastic processes \cite{Unser2012}. We also provide a summary of the notations in Table II.
\begin{table*}
\caption{Summary of notations}
\centering
\begin{tabular}{cll}
 \hline \\[-1ex]
Symbols & Description & Defining formula\\[2ex]
\hline\\[-1.5ex]
Innovation parameters:\\$V$ & L\'evy measure & $\displaystyle \int_\R \min(1,a^2)\  V(da) < \infty$ \\[1ex]
$v(a)$ & L\'evy density & $v(a)\ge0$ and $\displaystyle v(a) \dint a=V(da)$  \\[1ex]
$f(\omega)$ & L\'evy exponent & L\'evy-Khinchine formula\\[1ex]$p_A(a)$ & Poisson amplitude distribution & $p_A(a)\ge0$ and $\int_\R p_A(a) \dint a=1$  \\[1ex]
\hline\\[-1.5ex]
Stochastic differential equations: \\$\Lop$ & whitening operator & $\Lop s=w$: white noise\\[1ex]
$\rho_\Lop$& Green function& $\Lop\{\rho_\Lop\}=\delta$: Dirac impulse\\[1ex]
$\Identity$ & identity operator & 
\\[1ex]
$\Dop$ & derivative operator & $\Dop=\frac{\dint }{\dint t}$ \\[1ex]
$\Pop_\alpha$& first-order operator& $\Pop_\alpha=\Dop-\alpha \Identity$\\[1ex]
$N$ & order of differential system (number of poles) \\[1ex]
$n_0$ & order of unstability (number of imaginary poles) & $0 \le n_0 \le N$\\[1ex]
$\boldsymbol \alpha$ & vector of poles & $\boldsymbol \alpha =(\alpha_1,\dots,\alpha_N)$ \\[0ex]
$P_N(\s)=P_{\boldsymbol \alpha}(\s)$ & characteristic polynomial &  $\displaystyle P_N(\s)=\s^N+ \cdots+a_1\s+ a_0=\prod_{n=1}^N (\s-\alpha_n)$\\[1ex]
$\Pop_{\boldsymbol \alpha}=P_N(\Dop)$& $N$th-order differential operator& $\Pop_{\boldsymbol \alpha}=\Dop^N+\cdots+a_1\Dop+a_0\Identity=\Pop_{\alpha_1} \cdots \Pop_{\alpha_N}$\\[1ex]
$\hat \rho_\Lop(\omega)=\frac{1}{\hat L(\omega)}$ & rational transfer function & $\displaystyle \hat \rho_\Lop(\omega)=\frac{Q_M(j\omega)}{P_N(j\omega)}$\\[2ex]
\hline\\[-1.5ex]
Exponential B-splines:\\
$\Delta_{\alpha}$&first-order difference operator&$\Delta_{\alpha}f(t)=f(t)-e^{\alpha}f(t-1)$\\[1ex]
$\Delta_{\boldsymbol \alpha}$&$N$th-order difference operator&$\displaystyle \Delta_{\boldsymbol \alpha}f(t)=\Delta_{\boldsymbol \alpha_1}\cdots \Delta_{\alpha_N}f(t)=\sum_{n=0}^N d_{\boldsymbol \alpha}[k] f(t-k)$\\[1ex]
$D_{\boldsymbol \alpha}(z)$ & localization filter & $\displaystyle D_{\boldsymbol \alpha}(z)=\prod_{n=1}^N (1-e^{\alpha_n}z^{-1})=\sum_{n=0}^N d_{\boldsymbol \alpha}[k] z^{-k}$ \\[1ex]
$\beta_{\boldsymbol \alpha}(t)$ & exponential B-spline & $\displaystyle\beta_{\boldsymbol \alpha}(t)=\int_\R e^{j \omega t}\prod_{n=1}^N \left(\frac{1-e^{\alpha_n-j\omega} }{j\omega-\alpha_n}\right)\frac{\dint \omega}{2 \pi}$ \\[1ex]
$\beta_{\Lop}(t)$ & generalized B-spline & $\displaystyle\beta_{\Lop}(t)=\Delta_{\boldsymbol \alpha} \rho_\Lop(t)=\int_\R  \frac{D_{\boldsymbol \alpha}(e^{j \omega})}{\hat L(\omega)}e^{j \omega t} \frac{\dint \omega}{2 \pi}$ \\[2ex]
 \hline
\end{tabular}
\end{table*}
\subsection{Generalized innovation models}
\label{sec:innov}
The continuous-time stochastic processes $s(t)$ under consideration satisfy the general innovation model in Fig. 1. They correspond to the solution of the (linear) operator equation
\begin{equation}
\label{eq:operatoreq}
\Lop s=w,
\end{equation}
where the driving term $w$ is a continuous-domain white
noise process. The model has the ability to generate Gaussian processes, as well as a broad variety of sparse processes, depending upon the type
of excitation noise. 
The delicate aspect 
is that the underlying innovations $w$ do not admit a standard (pointwise) interpretation as functions of $t$ because they are highly singular. They can only be properly specified as distributions (a.k.a. generalized functions). Thus, the correct interpretation of (\ref{eq:operatoreq}) is in the ``weak" sense of distributions:
$$ \langle \varphi, \Lop s \rangle=\langle \varphi, w\rangle, \mbox{  for all  } \varphi \in \mathcal{S}$$
where the equality must hold true for any smooth and rapidly-decreasing test function $\varphi$ in Schwartz's class $\mathcal{S}$. The guiding principle is that, for any given $\varphi$, the scalar product (or linear functional) $\langle \varphi, w\rangle$ is a well-defined scalar random variable no matter how rough the actual innovation process $w$ is.

As for the class of admissible\footnote{A stochastic process is called white noise iff. it is stationary and independent at all points. In our framework, this is equivalent to requiring that the random observation variables $x_1=\langle \varphi_1, w\rangle$ and $x_2=\langle \varphi_2, w\rangle$ are: 1) identically-distributed whenever $\varphi_2(t)=\varphi_1(t-t_0)$ for any  $t_0 \inR$ (translated observations), and 2) independent whenever $\varphi_1 \times \varphi_2=0$ (observation windows with disjoint support).}
 input innovations, we have pointed out that 
each brand 
is uniquely characterized by a canonical infinitely divisible distribution $p_{\rm id}(x)$ (or, equivalently, a L\'evy exponent $f$) which specifies the PDF of its ``pixelated" observation (through a rectangular window) $x=\langle w, {\rm rect}(\cdot-t_0)\rangle$ which is i.i.d. and independent upon $t_0$ (stationarity).

The above innovation model is exploitable only if the whitening operator $\Lop$ has an inverse that is well-defined over an appropriate subset of $\mathcal{S}'$ (the space of tempered distributions). 
The equation is then solved formally as
\begin{align}
s&=\Lop^{-1}w \quad
\Leftrightarrow \quad \onetwocol{Ê}{\nonumber \\ &}\forall \varphi \in \mathcal{S}, \quad  \langle \varphi, s \rangle=\langle \varphi, \Lop^{-1}w \rangle=\langle \Lop^{-1\ast}\varphi, w \rangle
\label{eq:PDEinv}
\end{align}
where we are using a standard duality argument to move the action of the inverse operator (via its adjoint $\Lop^{-1\ast}$) onto the test function $\varphi$.
We have shown \cite[Theorem 3]{Unser2012} that a sufficient condition for this method of solution to yield a well-defined stochastic process $s$ is
\begin{eqnarray}
\label{eq:stable}
\forall \varphi \in \mathcal{S}, \quad\|\Lop^{-1\ast}\varphi\|_{L_p}<C \|\varphi\|_{L_p}
\end{eqnarray}
for some constant $C$ and $p\ge1$, which puts some mathematical constraints on the class of 
admissible operators and excitation noises. The implicit requirement is that the excitation noise is $p$-admissible, which is a condition imposed on its L\'evy exponent $f(\omega)=\log\hat p_{\rm id}(\omega)$  (cf. Definition \ref{def:padmis}, Section \ref{sec:charfunc}).

\subsection{$N$th-order stochastic differential equations}
\label{sec:solutionODE}
We have demonstrated that the above operator method could be deployed for finding the solutions of the complete class of linear stochastic differential equations of the form
\begin{eqnarray}
\label{eq:ODE}
 \sum_{n=1}^N a_n \Dop^n s=  \sum_{m=1}^M b_m \Dop^m w
\end{eqnarray}
with $N>M$, where $a_n$ and $b_m$ are arbitrary complex coefficients with the normalization constraint $a_N=1$, irrespective of any stability considerations.  The driving noise $w$, which constitutes the input of the system, is assumed to be white by default. The output $s(t)$ is our generalized stochastic process whose sample values are generally well-defined due to the smoothing effect of the inverse operator $\Lop^{-1}$.
The characteristic polynomial of the underlying $N$th-order system with Laplace variable $\zeta\in \C$ is
\begin{align}
P_N(\zeta)&=\zeta^N+a_{N-1}\zeta^{n-1}+\cdots+ a_0\onetwocol{ }{\nonumber \\&}=\prod_{n=1}^N (\zeta-\alpha_n)=P_{\boldsymbol \alpha}(\zeta),
\label{eq:pol}
\end{align}
and is also specifiable in term of its (complex) roots; these are collected in the vector of poles $\boldsymbol \alpha =(\alpha_1,\dots,\alpha_N)$ with the understanding that the notations $P_N(\zeta)$ and $P_{\boldsymbol \alpha}(\zeta)$ are equivalent. 

The linear system specified by (\ref{eq:ODE}) is causal-stable iff. all its poles are in the left complex half-plane. 
Under this classical assumption, its impulse response $\rho_\Lop(t)=\Lop^{-1}\{\delta\}(t)$
is exponentially decaying. It is obtained by taking the inverse Fourier transform of the rational transfer function
\begin{eqnarray}
\label{eq:transfer}
\hat \rho_\Lop(\omega)=\frac{Q_M(j\omega)}{P_N(j\omega)}=b_M\frac{ \prod_{m=1}^M (j\omega - \gamma_n)}{\prod_{n=1}^N (j\omega - \alpha_n) }=\frac{1}{\hat L(\omega)} ,
\end{eqnarray}
where $Q_M(\s)=b_M\s^M+b_{M-1}\s^{M-1}+\cdots+b_1\s +b_0$ is a polynomial of degree $M<N$.
The roots of $Q_M(\s)$ are the so-called zeros: $\boldsymbol \gamma =(\gamma_1,\dots,\gamma_M)$. 
The solution (output of the system) is then given by $s(t)=(\rho_\Lop \ast w)(t)$ and is stationary by construction (because of the shift-invariant filtering). When the excitation is Gaussian, one obtains the conventional continuous-time ARMA processes, but one can also generate a large variety of sparse counterparts of these processes by switching to appropriate types of non-Gaussian L\'evy innovations.

Remarkably, the proposed framework can also handle the unstable scenarios, the general rule being that each pole located on the imaginary axis  induces one degree of non-stationarity. 
Our extended formulation requires a special ordering of the poles where the $n_0$ purely-imaginary roots (if present) are coming last. This gets translated in the following  representation of the characteristic polynomial (\ref{eq:pol}): 
\begin{align}
P_{\boldsymbol \alpha}(j\omega)&=\left(\prod_{n=1}^{N-n_0}(j\omega - \alpha_n)\right)\,\left( \prod_{m=1}^{n_0}(j\omega - j \omega_m)\right)
\end{align}
with  $\alpha_{N-n_0+m}=j \omega_m$ and $\omega_m\inR$. It allows us to write the factorized version of the differential equation (\ref{eq:ODE}):
\begin{eqnarray}
\label{eq:ODEfac}
(\Pop_{\alpha_1} \cdots \Pop_{\alpha_{N-n_0}})(\Pop_{j\omega_1}\cdots \Pop_{j\omega_{n_0}})\{s\}=Q_M(\Dop)\{w\}
\end{eqnarray}
where $\Pop_{\alpha_n}=(\Dop-\alpha_n\Identity)$ is the operator counterpart of the Fourier multiplier $(j\omega-\alpha_n)$ and $Q_M(\Dop)=\sum_{m=1}^M b_m\Dop^m$. Each component $\Pop_{\alpha_n}$ with ${\rm Re}(\alpha_n)\ne0$ has a stable linear shift-invariant (LSI) inverse $\Pop_{\alpha_n}^{-1}$, which is either causal or anti-causal depending of the polarity of $\alpha_n$. The only delicate step in solving (\ref{eq:ODEfac}) is the inversion of the second operator factor on the left which is ill-posed. Our contribution has been to propose a stable inversion mechanism that makes use of some ``regularized" left inverse of $\Pop_{j\omega_0}$. The canonical solution is
\begin{eqnarray}
\label{eq:inv1}
\displaystyle \Iop_{\omega_0,\delta}f(t) =\int_{\R}   \hat{f}(\omega) \left(
   \frac{e^{j \omega t}-e^{j \omega_0t} }
       {j(\omega-\omega_0)} \right)
    \frac{\dint{\omega\;\;}}{2 \pi}
\end{eqnarray}
which, in accordance with (\ref{eq:genboundary}),  forces the output signal to vanish at $t=0$. 
This ultimately yields the global inverse operator 
\begin{eqnarray}
\label{eq:invODE}
\Lop^{-1}=\underbrace{\Iop_{\omega_{n_0}, \delta} \cdots \Iop_{\omega_{1}, \delta} }_{\mbox{shift-variant}}\; \underbrace{\Pop_{N-n_0}^{-1} \cdots \Pop_{\alpha_1}^{-1} Q_M(\Dop)}_{\mbox{ LSI part}},
\end{eqnarray}
to be substituted in (\ref{eq:PDEinv}); the latter imposes the $n_0$ boundary conditions on the output
\begin{eqnarray}
\label{eq:boundary}
\left\{
\begin{array}{rcl}
\left. s(0)\right.&=&0  \\
\left. (\Dop-j \omega_{n_0}\Identity)\{s\}(0)\right.&=&0\\
&\vdots&\\
\left.  (\Dop-j \omega_{2}\Identity)\cdots(\Dop-j \omega_{n_0}\Identity) \{s\}(0)\right.&=&0.
\end{array}
\right.
\end{eqnarray}
We have shown that this method of solution yields a generalized CARMA process $s=\Lop^{-1}w$ that is mathematically well-defined. Such processes will exhibit a $n_0$ degree of non-stationarity due to the lack of shift-invariance of
the elementary inverse operators $\Iop_{\omega_m,\delta}$. While the above inversion method is uniquely tied to the boundary conditions (\ref{eq:boundary}), it is not the only possible approach.
In the appendix, we show that one can impose other boundary conditions (in the form of $n_0$ generalized linear constraints: $\langle s, \varphi_m\rangle=0, \ {\tiny m=1,\dots,n_0}$), while retaining the required functional properties of the corresponding inverse operators $\Iop_{\omega_m,\varphi_m}$ and their adjoint.

The simplest example of unstable scenario is $\Dop s=w$, which corresponds to a single pole at the origin: $\alpha_1=j \omega_1=0$ and $N=n_0=1$. The  solution
$s(t)=\Iop_{0,\delta}w(t)=\int_{0}^t w(\tau)\dint \tau$, which enforces the boundary condition $s(0)=0$, perfectly maps into the L\'evy processes, although these are usually described quite differently
\cite{Bertoin1996,Sato1994}. The interest here is that we are constructing the L\'evy processes as the (unstable) limit of the non-Gaussian AR(1) family. We will see that this novel point of view facilitates the transposition of standard signal processing techniques to the non-stationary/non-Gaussian L\'evy setting, including the higher-order extensions of such processes.
\subsection{Characteristic functional}
\label{sec:charfunc}
Under the assumption that the whitening operator $\Lop$ admits an inverse that meets the stability condition (\ref{eq:stable}), we have shown that the generalized stochastic process $s(t)$ satisfying the innovation model (\ref{eq:operatoreq}) is completely and uniquely characterized
by its characteristic functional (cf. \cite[Theorem 3]{Unser2012}):
\begin{eqnarray}
\label{eq:gennoise}
\Form_s(\varphi)&=&\E\{e^{j \langle s, \varphi \rangle}\}\nonumber\\
&=&\exp\left( \int_{\R} f\big(\Lop^{-1\ast}\varphi(t)\big) \dint t\right)
\end{eqnarray}
under the constraint that the so-called L\'evy exponent $f(\omega)$ is $p$-admissible for the same $p$ as in (\ref{eq:stable}).
\begin{definition}
\label{def:padmis}
$f(\omega)$ is a $p$-admissible L\'evy exponent  for some $p>0$ iff. ({\em i}) it admits a L\'evy-Khinchine representation (cf. \cite[Eq. (8)]{Unser2012}) with some L\'evy triplet $(b_1,b_2,v(a))$, and ({\em ii}) $|f(\omega)|+|\omega| \ |f'(\omega)|< C |\omega|^p$. 
 \end{definition}
 

The powerful aspect of the formulation is that the functional $\Form_s(\varphi): \mathcal{S} \rightarrow \C$, which is the conceptual equivalent of an infinite-dimensional
characteristic function, condenses all the statistical information about the process.
The underlying principle is that the inverse operator $\Lop^{-1}$ (generalized shaping filter) specifies the covariance structure (or generalized spectrum) of the process $s$, while the L\'evy exponent $f(\omega)$ fully embodies the statistical properties of the innovation $w$. 

The classical choice of L\'evy exponent in \eqref{eq:gennoise} is $f_{\rm Gauss}(\omega)=-b_2 |\omega|^2 $ which results in the specification of the complete class of Gaussian processes. The remarkable aspect of the theory is that any other admissible choice induces a sparse behavior.
For instance, the generic L\'evy triplet $(0,0,\lambda p_A(a))$ where $\lambda>0$ and $p_A(a)$ is a valid pdf 
results in the definition of the extended class of generalized Poisson processes with $f_{\rm Poisson}(\omega; \lambda, p_A)=\lambda \int_{\R} \big(e^{j a \omega} - 1\big)\, p_A(a) \dint a$ \cite{Unser2011}. The latter is $p$-admissible with $p=1$ (provided that $\int_\R |a| p_A(a)\dint a<\infty$) and/or $p=2$ when $p_A(a)$ is symmetric.
The corresponding innovation is a sequence of randomly scattered Dirac impulses with Poisson parameter $\lambda>0$ (average number of singularities per unit time) and amplitude distribution $p_A(a)$. Also included in the framework are the symmetric-alpha-stable (S$\alpha$S) processes (with $f_{\alpha}(\omega)=-b_\alpha |\omega|^\alpha)$, which, for $0<\alpha<2$, have the intriguing property that their second-order moments are unbounded (heavy tail behavior)\cite{Samorodnitsky1994}. For a more details, refer to \cite[Sections III.C-D]{Unser2012}.

\section{Generalized increment process}
Since $\Lop^{-1}$ is typically an integral operator, its effect on $s=\Lop^{-1}w$ is to induce long range dependencies. These need to be suppresses
if one wishes to obtain a sparse signal representation.
The first approach investigated in \cite{Unser2012} is to apply a wavelet transform where the wavelets act as multiresolution versions of the whitening operator $\Lop$. While the decoupling effect of such an analysis is adequate within a given scale, we have seen that it not quite as favorable between scales because of the overlap of the underlying smoothing kernels.

In principle, we could get back to the innovation by simply applying $\Lop$ to $s$. Unfortunately, this is not feasible in practice since we only have the samples 
of the process available. The best computational strategy is to apply a discrete version of the operator $\Lop$ which we shall denote by $\Lop_{\rm d}$.
The main point that we shall make in this section is that applying $\Lop_{\rm d}$ to $s$ is equivalent to smoothing the innovation with a localized kernel $\beta_\Lop$ (generalized B-spline):
\begin{align}
\Lop_{\rm d}s(t)=(\beta_\Lop \ast w)(t)
\end{align}
where $\beta_\Lop=\Lop_{\rm d}\Lop^{-1}\delta=\Lop_{\rm d}\rho_\Lop$. 
To get the best decoupling effect, we need to select $\Lop_{\rm d}$ such that $\beta_\Lop$ is most localized---ideally, compactly supported. The good news is that we can rely on spline mathematics to identify the shortest solution. As far as statistics are concerned, it is also useful to recall that the innovation process $w$ is completely and uniquely specified its characteristic form
\begin{align}
\label{eq:levyfunctional}
\Form_w(\varphi)=\E\{e^{\langle w, \varphi\rangle}\}=\exp\left(\int_\R f\big(\varphi(t)\big)\dint t\right)
\end{align}
and hence by its L\'evy exponent $f: \R\to\C$ which is such that $f(0)=0$.

\subsection{Exponential B-splines and finite difference operators}
\label{sec:Bspline}
The foundation of exponential spline calculus is that we can always factor an $N$th-order differential operator into a cascade of first-order operators $\Pop_{\alpha_n}=(\Dop-\alpha_n\Identity)$
where the $\alpha_n$ (complex poles) are the roots of the characteristic polynomial; i.e.,
\begin{eqnarray*}
P_N(\Dop)&=&\Dop^N+a_{N-1}\Dop^{N-1}+\cdots+a_1\Dop+a_0 \Identity \\
&=&\Pop_{\alpha_N}\cdots \Pop_{\alpha_1}=\Pop_{(\alpha_1,\dots,\alpha_N)}
\end{eqnarray*}
where the right-hand side concatenated operator notation is self-explanatory. This allows us to express the Green function of $\Pop_{\boldsymbol \alpha}$ with pole vector ${\boldsymbol \alpha}=(\alpha_1,\dots,\alpha_N)$ as the convolution of the Green functions of its elementary constituents
\begin{eqnarray}
\rho_{\boldsymbol \alpha}(t)=(\rho_{\alpha_1} \ast \rho_{\alpha_2} \cdots \ast \rho_{\alpha_N})(t)
\end{eqnarray}
with 
\begin{eqnarray}
\label{eq:rhoalpha}
\rho_{\alpha}(t)=\left\{
\begin{array}{ll}
 \One_+(t) e^{\alpha t} & \text{if  ${\rm Re}(\alpha)\le 0$} \\
- \One_+(-t) e^{\alpha t} & \text{otherwise.} 
\end{array} \right.
\end{eqnarray}
The so-defined Green function $\rho_{\boldsymbol \alpha}(t)$ is necessarily of slow growth; it specifies the impulse response
of the LSI inverse operator $\Pop_{\boldsymbol \alpha}^{-1}$, which is well-defined over $\mathcal{S}$,
\begin{eqnarray*}
\Pop_{\boldsymbol \alpha}^{-1} \varphi(t) = (\rho_{\boldsymbol \alpha} \ast \varphi)(t),
\end{eqnarray*}
but not necessarily bounded (when some of the poles are purely imaginary).

Next, we observe that by applying the finite difference operator
\begin{eqnarray*}
\Delta_{\alpha}f(t)=
f(t)-e^{\alpha}f(t-1)
\end{eqnarray*}
to the function $\rho_{\alpha}(t)$, we are able to construct a compactly-supported function: the first-order exponential B-spline with parameter $\alpha$
\begin{eqnarray*}
\beta_{\alpha}(t)=\Delta_\alpha \rho_{\alpha}(t)=\left\{
\begin{array}{ll}
\One_{[0,1)}(t)e^{\alpha t} & \mbox{if  } {\rm Re}(\alpha)\le 0 \\
\One_{[0,1)}(t) e^{\alpha (t-1)} & \mbox{else}. 
\end{array} \right.
\end{eqnarray*}
The generalization of this scheme yields the $N$th-order B-spline with parameter vector ${\boldsymbol \alpha}=(\alpha_1,\dots,\alpha_N)$
\begin{eqnarray}
\beta_{\boldsymbol \alpha}(t)=\Delta_{{\boldsymbol \alpha}}\rho_{\boldsymbol \alpha}(t)=(\beta_{\alpha_1} \ast \beta_{\alpha_2} \cdots \ast \beta_{\alpha_N})(t).
\label{eq:BsplineN}
\end{eqnarray}
These functions have the following properties (cf \cite{Unser2005}):
\begin{itemize}
\item They are smooth and well-localized: compactly supported in $[0,N]$, bounded, and H\"older continuous of order $N-1$.
\item They are piecewise-exponential with joining points at the integer and a maximal degree of smoothness (spline property). For ${\boldsymbol \alpha}=(0, \dots,0)$, one recovers Schoenberg's classical polynomial B-splines of degree $N-1$ \cite{Schoenberg1946,Schoenberg1973}.
\item They are the shortest elementary constituents of splines: the functions $\{\beta_{\boldsymbol \alpha}(t-n)\}_{n\inZ}$ forms a Riesz basis of the corresponding family of exponential splines with knots at the integers. 
\end{itemize}
The crucial formula for our purpose is the equivalent operator interpretation of the B-spline formula (\ref{eq:BsplineN}):
\begin{eqnarray}
\Delta_{\boldsymbol \alpha} \Pop_{\boldsymbol \alpha}^{-1} \varphi = \Delta_\alpha\rho_{\boldsymbol \alpha}\ast \varphi
= \beta_{\boldsymbol \alpha} \ast \varphi, \label{eq:splinelocal}
\end{eqnarray}
which we will now put to good use in order to partially undo the effect of the inverse operator \eqref{eq:invODE}, or any variant thereof that imposes other linear boundary conditions.

\begin{theorem}
Let $\{\Iop^\ast_{\omega_m,\varphi_m}\}_{m=1}^{n_0}$ with $\omega_m \in \R$ be a
series of generalized (adjoint) inverse operators of the type defined by \eqref{eq:invop*} and let
$\{\Delta^\ast_{j\omega_m}\}_{m=1}^{n_0}$ be some corresponding adjoint
localization operators with $\Delta^\ast_{j\omega_m}\varphi(t)=\varphi(t)-e^{j
\omega_m}\varphi(t+1)$. Then, for all $\varphi\in\Spc S$,
\begin{eqnarray*}
 \Iop^\ast_{\omega_{1}, \varphi_1} \cdots \Iop^\ast_{\omega_{n_0}, \varphi_{n_0}} \Delta^\ast_{j\omega_{n_0}}\cdots \Delta^\ast_{j\omega_1} \varphi = \beta^\vee_{(j \omega_1,\dots,j\omega_{n_0})} \ast \varphi \\
\Delta_{j\omega_1}\cdots \Delta_{j\omega_{n_0}}  \Iop_{\omega_{n_0},
\varphi_{n_0}} \cdots \Iop_{\omega_{1}, \varphi_{1}}  \varphi = \beta_{(j
\omega_1,\dots,j\omega_{n_0})} \ast \varphi
\end{eqnarray*}
where $\beta_{(j \omega_1,\dots,j\omega_{n_0})}$ an exponential B-spline kernel as defined by (\ref{eq:BsplineN}). Since the latter is bounded and compactly-supported, the resulting convolution operators are BIBO-stable and $\mathcal{S}$-continuous.
\end{theorem}
\begin{proof} First, we observe that $\widehat{\Delta^\ast_{j\omega_m}f}(\omega)= (1-e^{j\omega_m}
e^{j\omega})\hat f(\omega)$.
Using Definition $\eqref{eq:invop*}$, we then evaluate the Fourier transform of $g(t)=\Iop^\ast_{\omega_{m}, \varphi_{m}}\Delta^\ast_{j\omega_m}f(t)$ as
\onetwocol{
\begin{align*}
\hat g(\omega)&=
   \frac{(1-e^{j\omega_m}e^{j\omega})\hat{f}(\omega)-(\overbrace{1-e^{j\omega_m}e^{-j\omega_m}}^{=0})\hat f(-\omega_m) \frac{\hat \varphi_m(\omega)} 
   {\hat \varphi_m(-\omega_m)}}
       {-j(\omega+\omega_m)}  \\
       &=\hat{f}(\omega) \left(
   \frac{1-e^{j\omega_m+j\omega} }
       {-j\omega-j\omega_m} \right),
\end{align*}
}{ \begin{align*}
\hat g(\omega)&=
   \frac{(1-e^{j\omega_m}e^{j\omega})\hat{f}(\omega)}
       {-j(\omega+\omega_m)}  \\
         &\hspace*{3ex}-\frac{(\overbrace{1-e^{j\omega_m}e^{-j\omega_m}}^{=0})\hat f(-\omega_m) e^{-j (\omega+\omega_m)t_m} }      {-j(\omega+\omega_m)}  \\
       &=\hat{f}(\omega) \left(
   \frac{1-e^{j\omega_m+j\omega} }
       {-j\omega-j\omega_m} \right),
\end{align*}
}
where we identify the right-hand side factor as $\hat
\beta_{j\omega_m}(-\omega)$ where $\hat
\beta_{\alpha}(\omega)=\frac{1-e^{\alpha-j\omega}}{j\omega-\alpha}$
is the Fourier transform of the first-order exponential B-spline
with parameter $\alpha$. This proves that $\Iop^\ast_{\omega_{m},
\varphi_{m}}\Delta^\ast_{j\omega_m}f=\beta_{j\omega_m}^\vee \ast f$ for
any $\omega_m,\varphi_m \inR$. Using the property that the order of
application of stable convolution operators such as
$\Delta^\ast_{j\omega_m}$ can be changed (commutativity), we start
with $\Iop^\ast_{\omega_{n_0}, \varphi_{n_0}}\Delta^\ast_{j\omega_{n_0}}f$ and
progressively work our way outwards to show that
$\Iop^\ast_{\omega_{1}, \varphi_1} \cdots \Iop^\ast_{\omega_{n_0}, \varphi_{n_0}}
\Delta^\ast_{j\omega_{n_0}}\cdots \Delta^\ast_{j\omega_1} \varphi =
\beta_{j \omega_1}^\vee \ast \cdots \ast \beta_{j \omega_{n_0}}^\vee
\ast \varphi$, which, thanks to (\ref{eq:BsplineN}), yields the
desired result. The second formula is established in the same way.
\end{proof}
The interpretation of the second relation is that the difference operators $\Delta_{j\omega_n}$ annihilate the sinusoidal components that are in the null space of $(\Dop-j \omega_n\Iop)$ so that the effect of $\Iop_{\omega_{m}, t_m}$ becomes indistinguishable from that of the non-regularized shift-invariant inverse $\Iop_{\omega_{m}}$. 
By combining this result with (\ref{eq:splinelocal}), we obtain a stable LSI substitute for the original inverse operator with the added benefit of a much better localization.
\begin{corollary}
\label{th:localization}
Let $\Lop^{-1}$ be the $N$th-order (not necessarily shift-invariant) inverse operator specified by (\ref{eq:invODE}).  Then,
\begin{eqnarray*}
\Lop^{-1\ast} \Delta^\ast_{\boldsymbol \alpha} \varphi =\beta_{\Lop}^\vee \ast \varphi \\
\Delta_{\boldsymbol \alpha} \Lop^{-1} \varphi = \beta_{\Lop} \ast \varphi,
\end{eqnarray*}
where $\beta^\vee_{\Lop}(t)=\beta_{\Lop}(-t)$ and $\beta_{\Lop}$ is the generalized B-spline kernel
\begin{eqnarray}
\label{eq:Lspline}
\beta_{\Lop}(t)=Q_M(\Dop)\beta_{\boldsymbol \alpha}(t)=\sum_{m=1}^M b_m\Dop^m\beta_{\boldsymbol \alpha}(t).
\end{eqnarray}
The latter is a linear combination of derivatives of the $N$th-order exponential B-spline $\beta_{\boldsymbol \alpha}(t)$ with parameter vector ${\boldsymbol \alpha}=(\alpha_1,\dots,\alpha_{N})$, and is therefore compactly-supported over the time-interval $[0,N]$.
\end{corollary}

The intuition behind this result is that we are localizing the system's response by canceling the poles of its frequency response; i.e., a pole at $j\omega=\alpha_n$  is neutralized by a corresponding zero of $1-e^{\alpha_n-j \omega}$ (the frequency response of $\Delta_{\alpha_n}$).

\subsection{Generalized increments and decoupling of sparse processes}
\label{Sec:gincr}
We shall now see that the application of the $N$th-order difference operator $\Delta_{\boldsymbol \alpha}=\Delta_{\alpha_1}\cdots\Delta_{\alpha_N}$ has the ability to partially decouple $s$.  This results in the natural extension of the classical notion of increments for Brownian motion and L\'evy processes (cf. \cite[Section VI.B]{Unser2012}).

\begin{proposition} [Generalized increment processes]
\label{Prop:increment}
Let $s$ be a generalized stochastic process whose characteristic form is $\Form_s(\varphi)=\Form_w(\Lop^{-1\ast}\varphi)$ where $\Form_w$ and
$\Lop^{-1}$ are specified by 
\eqref{eq:levyfunctional} and \eqref{eq:invODE}, respectively (differential system of order $N$ with pole vector $\boldsymbol \alpha$ and driving operator $Q_M(\Dop)=\sum_{m=1}^M b_m \Dop^m$). 
The corresponding generalized increment process
$$u(t)=\Delta_{\boldsymbol \alpha} s(t)$$
is well-defined and stationary (irrespective of any stability consideration). Its characteristic form is 
given by 
$\Form_{u}(\varphi)=\Form_{w}(\beta_{\Lop}^\vee \ast \varphi)
$
where $\beta_{\Lop}$ is the generalized B-spline kernel defined by (\ref{eq:Lspline}).
\end{proposition}

The result is also valid for all the variants of $\Lop^{-1\ast}$ described in the appendix, irrespective of the actual choice of boundary conditions (cf. Eqs.\
 \eqref{eq:ginvadj} and \eqref{eq:boundarygen}), since $\Delta_{\boldsymbol \alpha}$ removes the signal components in the null space of $\Lop$.
\begin{proof}: Corollary \ref{th:localization} implies that $\Delta_{\boldsymbol \alpha}\Lop^{-1}w=\beta_{\Lop}\ast w$. Since the convolution with the compactly-supported kernel $\beta_{\Lop}$ defines a continuous LSI operator on $\mathcal{S}$, we can  invoke \cite[Proposition 3]{Unser2012} with $\rho=\beta_{\Lop}$, which yields the desired result.
\end{proof}

Since the generalized B-spline $\beta_\Lop$ is H\"older-continuous of order $N-M-1$, the above characterization allows us to infer that 
the two processes $u$ and $s$ are $(N-M-2)$ times differentiable in the classical sense. 
In fact, the processes are well-defined pointwise as soon as $N>M$, which is the minimum requirement for continuity in the mean-square sense \cite{Adler1981}.
The other direct implication is that the samples of the generalized increment process, $u(t_1)$ and $u(t_2)$, are independent as soon as $|t_1-t_2|> N$ (due to the finite support property of the exponential B-spline $\beta_{\Lop}$). This means that working with the increment process $u(t)$ has the remarkable feature
of completely suppressing long-range dependencies.

\begin{property}[Reduction of correlation distances]
\label{Prop:correlation_distance}
Let $s$ be a generalized stochastic process whose characteristic form is $\Form_s(\varphi)=\Form_w(\Lop^{-1\ast}\varphi)$ where $\Form_w$ is a white noise functional \eqref{eq:levyfunctional} and where
$\Lop^{-1\ast}$ is given by (\ref{eq:ginvadj}) (differential system of order $N$ with pole vector $\boldsymbol \alpha$ and driving operator $Q_M(\Dop)=\sum_{m=1}^M b_m \Dop^m$). Then, the correlation form of 
$u(t)=\Delta_{\boldsymbol \alpha} s(t)$
can be written as $$
\Corr_{u}(\varphi_1, \varphi_2)= \sigma^2_0 \;\langle \beta_{\Lop}^\vee \ast \varphi_1,\overline{\beta_{\Lop}}^\vee \ast \varphi_2\rangle,
$$
where $\beta_{\Lop}$ is the generalized B-spline defined by (\ref{eq:Lspline}). The
corresponding covariance function is
\begin{align*}
R_{u}(t_1,t_2)&=\E\left\{\Delta_{\boldsymbol \alpha}s(t_1) \cdot  \overline{\Delta_{\boldsymbol \alpha}s(t_2)}\right\} \onetwocol{}{\\ &}
= \sigma^2_0 \; 
\left(\overline{\beta_{\Lop}} \ast \beta_{\Lop}^\vee\right)
(t_2-t_1) 
\end{align*}
which vanishes for $(t_2-t_1) \notin [-N,N]$.
\end{property}
The above result is universal in the sense that it does not distinguish between the stable and unstable cases; it can handle $N$th-order systems in full generality.

\section{Connection with discrete-time stochastic processes}
We will now show that there is an elegant connection between the continuous-time and discrete-time formulations of stochastic processes which is analogous to the connection that can be drawn between the corresponding deterministic linear system theories \cite{Unser2005d,Unser2005}.
The story in a nutshell is as follows: continuous-time processes are ruled by differential equations, while their discrete counterparts are solutions of difference equations.
The equations and correlation structures are linked functionally through some generalized compactly-supported B-splines. The use of these B-splines also greatly facilitates the transposition of the methods of solution from one domain to the other.
\subsection{Discrete-domain notations}
Discrete  processes and sequences are indexed using square brackets (e.g, $s[k]$, $h[k]$) to differentiate them from their continuous counterparts (e.g., $s(t)$ and $h(t)$).
A sequence $h[k]$ of slow growth (i.e., $h[k]$ does not grow faster at infinity than a polynomial of $k$) is characterized by its $z$-transform $H(e^{j \omega})=\sum_{k \inZ} h[k]z^{-k}$, which yields the discrete-time Fourier transform for $z=e^{j \omega}$.
If $h[k]=\left.h(t)\right|_{t=k}$ is the sampled version of the continuous function $h(t)$ with sufficient decay, then one can relate their discrete and continuous-time Fourier transforms using Poisson's summation formula:
$H(e^{j \omega})=\sum_{n \inZ} \hat h(\omega + 2 \pi n)$.

The localization operator $\Delta_{\boldsymbol \alpha}$ in Section \ref{Prop:increment} is transferable to the discrete domain; its discrete impulse response, denoted by $d_{\boldsymbol \alpha}[k]$,
is the inverse Fourier transform of $D_{\boldsymbol \alpha}(e^{j \omega})=\prod_{n=1}^N (1-e^{\alpha_n-j \omega})$, which coincides with the frequency response of the continuous-domain operator. The corresponding discrete notation is $\Delta_{\boldsymbol \alpha}s[k]=\left(d_{\boldsymbol \alpha} \ast s\right)[k]=\sum_{n \inZ} d_{\boldsymbol \alpha}[n]Ês[k-n]$, where the use of the square brackets indicates that the convolution operation is discrete.
\subsection{Sampled processes}
Here we will consider (ordinary) discrete stochastic processes that are sampled versions of the generalized ones:
$$s[k]=\langle s, \delta(\cdot -k)\rangle=\left.s(t)\right|_{t=k}$$
where $s(t)$ is the continuous-time solution of (\ref{eq:ODE}).
It should be clear now that the statistics of this discrete process are completely specified by $\Form_s(\varphi)$ in (\ref{eq:gennoise}). For instance, we may obtain its $K$th-order characteristic function 
$\E\{e^{j \langle {\bf s}, \bw \rangle}\}=\int_{\R^K} p_s({\bf s})e^{j \langle {\bf s}, \bw \rangle}\dint {\bf s}$ with ${\bf s}=(s[k], s[k-1], \dots, s[k-K+1])$ and $\bw=(\omega_1,\dots,\omega_K)$ for any finite $K$ by substituting 
$\varphi=\omega_1\delta(\cdot)+\omega_2\delta(\cdot-1)+ \cdots + \omega_K \delta(\cdot-K+1)$ in the characteristic form. Likewise, one can determine its correlation sequence by 
 sampling the continuous-time correlation function (as given by \cite[Property 1]{Unser2012}) on the integer grid: $R_s[k_1,k_2]=\left.R_s(t_1,t_2)\right|_{t_1=k_1, t_2=k_2}$.
 
Our objective is now to relate these quantities to the
Hermitian-symmetric Green function of the operator $\overline{\Lop} \Lop^\ast$. 
The latter, which is the distributional solution of $\overline{\Lop} \Lop^\ast \rho_{\overline{\Lop} \Lop^\ast }=\delta$, can formally be specified as  
\begin{eqnarray}
\label{eq:GreenLL}
\rho_{\overline{\Lop} \Lop^\ast}(t)=\int_{-\infty}^{±\infty} \frac{e^{j \omega t}}{|\hat L(-\omega)|^2}\frac{\dint \omega}{2 \pi}
\end{eqnarray}
where $\hat L(\omega)$ (resp., $|\hat L(-\omega)|^2$) is the transfer function of the LSI whitening operator $\Lop$ (resp., $\overline{\Lop} \Lop^\ast$). Note that in the singular case, the above integral has to be interpreted as a finite part (F.P.) integral in the sense of Hadamard.
In the event where $\Lop^{-1}$ is LSI BIBO-stable with impulse response $\rho_\Lop$, then $\rho_{\overline{\Lop} \Lop^\ast}=\overline{\rho_\Lop} \ast \rho_\Lop^\vee$.
However, in the unstable case, the latter convolution product is generally undefined; e.g., $\rho_{\Dop\Dop^\ast}(t)=\Fourier^{-1}\left\{\frac{1}{|\omega|^2} \right\}(t)=-\frac{1}{2}|t|\ne (u \ast u^\vee)(t)$ where the right-hand side expression is not converging anywhere. Next, we make the link with exponential splines by expressing the Green function as a weighted sum of augmented B-splines:
\begin{eqnarray}
\label{eq:Greenauto}
\rho_{\overline{\Lop} \Lop^\ast}(t)=\sum_{k \inZ} q_{\boldsymbol \alpha}[k] \beta_{\overline{\Lop} \Lop^\ast} (t-k) 
\end{eqnarray}
where $q_{\boldsymbol \alpha}[k]$ is the Hermitian-symmetric sequence whose discrete-time Fourier transform is
$$
Q_{\boldsymbol \alpha}(z)=\frac{1}{|D_{\boldsymbol \alpha}(e^{-j \omega})|^2}=\frac{1}{|\hat \Delta_{\boldsymbol \alpha}(-\omega)|^2}.
$$
The augmented B-spline kernel $\beta_{\overline{\Lop} \Lop^\ast}$ is given by
\begin{eqnarray}
\label{eq:augBspline}
\beta_{\overline{\Lop} \Lop^\ast}=\overline{\beta_{\Lop}} \ast \beta_{\Lop}^\vee=\overline{\Delta_{\boldsymbol \alpha}}\Delta^\ast_{\boldsymbol \alpha}\rho_{\overline{\Lop} \Lop^\ast}
\end{eqnarray}
where $\beta_{\Lop}$ is defined by (\ref{eq:Lspline}). Establishing (\ref{eq:Greenauto}) is a simple matter of factorization in the Fourier domain.
What is not so obvious at first sight is that the above entities are always well-defined, irrespective of any stability considerations. The generalized exponential B-spline $\beta_{\overline{\Lop} \Lop^\ast}(t)$, in particular, is compactly-supported in $[-N,+N]$ and guaranteed to yield a stable expansion (Riesz basis property) \cite[Theorem 1]{Unser2005d}.
$\rho_{\overline{\Lop} \Lop^\ast}(t)$ and $q_{\boldsymbol \alpha}[k]$, on the other hand,  are both infinitely-supported;  they are either exponentially-decaying (stable scenario with ${\rm Re}(\alpha_n)\ne0$) or, at worst, of slow (polynomial) growth when $n_0>0$.  Our final theoretical tool  is a corresponding exponential spline interpolation mechanism.

\begin{property}[Exponential spline interpolation] 
\label{Prop: interpolation}
Let $f(t)$ be a function (at most of slow growth) that is included in the exponential spline space $V_{\overline{\Lop} \Lop^\ast}={\rm span}\{\beta_{\overline{\Lop} \Lop^\ast}(t-k)\}_{k \inZ} \subset \mathcal{S}'$ where $\beta_{\overline{\Lop} \Lop^\ast}$ is specified by (\ref{eq:augBspline}) and compactly-supported in $[-N,N]$. Then,
$$
f(t)=\sum_{k \inZ} f(k) \varphi_{\rm int} (t-k)
$$
where $\varphi_{\rm int}(x)\in V_{\overline{\Lop} \Lop^\ast}$ is an exponentially-decaying interpolation function whose Fourier-domain expression is
\begin{eqnarray*}
\hat \varphi_{\rm int}(\omega) =\frac{\hat \beta_{\overline{\Lop} \Lop^\ast}(\omega)}{B_{\Lop}(e^{j \omega})}
\end{eqnarray*}
with
\begin{eqnarray}
\label{eq:DiscreteBsplinekern}
B_{\Lop}(z)=\sum_{k=-N}^N \beta_{\overline{\Lop} \Lop^\ast}(k) z^{-k}.
\end{eqnarray}
\end{property}
\begin{proof} The statement $f(t) \in V_{\overline{\Lop} \Lop^\ast}$ is equivalent to $f(t)=\sum_{k \inZ} c[k] \beta_{\overline{\Lop} \Lop^\ast}(t-k)$ where $c[k]$ is a sequence of (possibly slowly-growing) B-spline coefficients. By sampling this expression at the integers and taking the $z$-transform, we obtain $F(z)=\sum_{k \inZ}f(k)z^{-k}=C(z) B_{\Lop}(z)$ so that
$C(z)=F(z)/B_{\Lop}(z)$. The time-domain interpretation is that $c[k]= (h_{\rm int} \ast f) [k]$ where $h_{\rm int}$ is the impulse response of the (inverse) digital filter whose frequency response is $H_{\rm int}(e^{j \omega})=1/B_{\Lop}(e^{j \omega})$. Whenever the purely-imaginary poles of $1/\hat L(\omega)$ are such that $j \omega_n-j\omega_m\ne j 2 \pi k$ for any $n\ne m$ and $k\ne0$,  then $\beta_\Lop$ generates a Riesz basis \cite[Theorem 1]{Unser2005d}, which is equivalent to $0<A<B_{\Lop}(e^{j \omega})<B$ for any $\omega\inR$ ($A$ and $B$ are the lower and upper Riesz bounds of the B-spline basis). Therefore, by Wiener's lemma, we have the guarantee that the sequence $h_{\rm int}$ is well-defined ($h_{\rm int} \in \ell_1$) and exponentially-decreasing because $\beta_{\overline{\Lop} \Lop^\ast}(k)$ is compactly-supported. This leads to the conclusion that
$f(t)=\sum_{k \inZ} (h_{\rm int} \ast f)[k] \beta_{\overline{\Lop} \Lop^\ast}(t-k)=\sum_{k \inZ} f[k] \varphi_{\rm int} (t-k)$
where 
$\varphi_{\rm int} (t)=\sum_{k \inZ} h_{\rm int}[k] \beta_{\overline{\Lop} \Lop^\ast}(t-k)$
is exponentially-decaying as well. The Fourier transform of this last expression is $\hat \varphi_{\rm int} (\omega)=H_{\rm int}(e^{j \omega}) \hat \beta_{\overline{\Lop} \Lop^\ast}(\omega)$.
\end{proof}

We are now ready to uncover the relation between the second-order statistical characterizations of the continuous-time and discrete-time versions of our stochastic processes. For simplicity, we focus on the stationary case where the underlying $N$th-order system is stable (cf. \cite[Proposition 3]{Unser2012}). 

\begin{property}[Conversion from discrete to continuous] Let $s$ be a generalized (Gaussian or non-Gaussian) stationary process that satisfies the $N$th-order stochastic differential equation (\ref{eq:ODE}) with a white noise excitation. Then, the correlation functions of the continuous-time and discrete-time (e.g., sampled) instances of the process are linked through the  interpolation formula
\begin{eqnarray*}
r_s(t)=\E\{s(t')\cdot \overline{s(t'+t)}\}
&=&\sum_{k \inZ}r_s[k]\varphi_{\rm int} (t-k)
\end{eqnarray*}
where $\varphi_{\rm int}(x)$ is specified in Property \ref{Prop: interpolation} and $r_s[k]=\left.r_s(t)\right|_{t=k}$. The Fourier-domain counterpart of this expression provides the exact link 
between the continuous and discrete-domain power spectra of the process:
\begin{eqnarray*}
\Phi_s(\omega)=
\hat \varphi_{\rm int}(\omega) \Phi_{s}(e^{j\omega}).
\end{eqnarray*}

\end{property}

{\em Remark on notation}: While we are using a common symbol to denote the continuous and discrete autocorrelation (resp., power spectrum) of $s$, we are relying on the index variables to distinguish between the two settings. Specifically, $\Phi_s(\omega)=\Fourier\{r_s(t)\}(\omega)$ is the Fourier transform of the continuous-time autocorrelation function $r_s(t)$, while $\Phi_s(z)$ is the $z$-transform of the discrete-time correlation sequence $r_s[k]$ (or, equivalently, the discrete-time Fourier transform if we set  $z=e^{j \omega}$).

\begin{proof} Since the discrete process is the sampled version of the continuous one, we have that
$r_s[k]=\left.r_s(t)\right|_{t=k}$, or equivalently, $\Phi_{s}(e^{j\omega})=\sum_{n \inZ}\Phi_s(\omega + 2 \pi n)$.
We also know that $r_s(t)=\sigma_0^2 \rho_{\overline{\Lop} \Lop^\ast}(t)$ and $\Phi_s(\omega)=\frac{\sigma_0^2}{|\hat L(-\omega)|^2}$, as a direct consequence of the innovation model. Putting these elements together, we find that
\begin{eqnarray}
\label{eq:intpol2}
\frac{ \Phi_{s}(\omega)Ê}{ \Phi_{s}(e^{j\omega})Ê}=
\frac{\displaystyle \sum_{n \inZ}|\hat L(-\omega + 2 \pi n)|^2}{|\hat L(-\omega)|^2},
\end{eqnarray}
where $\hat L(\omega)$ is the frequency response of the whitening filter specified by the reciprocal of (\ref{eq:transfer}). 
We then use the B-spline connection to show the above ratio is well-defined and equal to $\hat \varphi_{\rm int}(\omega)$. To that end, we consider the Fourier-domain version of (\ref{eq:augBspline}) \begin{eqnarray*}
\hat \beta_{\overline{\Lop} \Lop^\ast}(\omega)
=\frac{|D_{\boldsymbol \alpha}(e^{-j \omega})|^2}{ |\hat L(-\omega)|^2}.\end{eqnarray*}
together with its periodized counterpart $\sum_{n \inZ} \hat \beta_{\overline{\Lop} \Lop^\ast}(\omega+ 2 \pi n)=\frac{|D_{\boldsymbol \alpha}(e^{-j \omega})|^2}{ \sum_{n \inZ}  |\hat L(-\omega+ 2 \pi n)|^2}=B_\Lop(e^{j \omega})$ (by Poisson's summation formula and the $2\pi$-periodicity of $D_{\boldsymbol \alpha}(e^{j \omega})$). It now suffices to express the right-hand side of (\ref{eq:intpol2}) as the ratio of these two entities, which yields the desired result. The main point of this manipulation is that $B_\Lop(e^{j \omega})$ is guaranteed to be non-vanishing (due to suitable pole-zero cancellations), while it is not necessarily so for the denominator of (\ref{eq:intpol2}).
\end{proof}

\subsection{Discrete increment process}
The important point that has been brought out by the above analyses is that the present class of discrete (or continuous-time) processes exhibit long-range dependencies due to the infinite support of their autocorrelation function. This behavior is further exacerbated in the non-stationary case where the (asymptotic) decay is linear at best.
Fortunately, we have seen that there is a simple way to obtain a much better conditioned signal by applying the localization operator $\Delta_{\boldsymbol \alpha}$ (cf. Proposition \ref{Prop:increment}). The good news is that this concept is directly transposable to the discrete domain as well, and that it substantially simplifies the statistical characterization of such signals, irrespective of any stability considerations.

Specifically, the discrete generalized increment process of $s[k]$ is defined as:
\begin{eqnarray}
\label{eq:discrete_Increment}
u[n]=\left.\Delta_{\boldsymbol \alpha}s(t)\right|_{t=n}=\sum_{m=0}^N d_{\boldsymbol \alpha}[m] s[n-m]
\end{eqnarray}
where $s(t)$ is a generalized $N$th-order stochastic process with whitening operator $\Lop$ and pole vector ${\boldsymbol \alpha}=(\alpha_1,\dots,\alpha_N)$;
the discrete AR-type filtering coefficients on the right hand side of (\ref{eq:discrete_Increment}) are given by
\begin{eqnarray}
\label{eq:Dlocal}
D_{\boldsymbol \alpha}(z)=\sum_{m=0}^N d_{\boldsymbol \alpha}[m] z^{-m}=\prod_{n=1}^N (1-e^{\alpha_n}z^{-1}).
\end{eqnarray}
%
%
\begin{property}[Characterization of discrete increment process]
\label{Prop:discreteincrement}
Let $u[k]$ be the discrete increment process associated with a (possibly non-stationary) $N$th-order generalized process whose characteristic functional $\Form_s(\varphi)$ is given by (\ref{eq:gennoise}) where
$\Lop^{-1\ast}$ is the adjoint of $\Lop^{-1}$ specified by (\ref{eq:invODE}) (see also \cite[Eq. (25)]{Unser2012}).
Then, $u[k]$ is stationary with an $N$th-order of dependency: $p_{u}\left(u[k]\left| \{u[k-m]\}_{m \inZ^+}\right)\right.=p_{u}(u[k]\left|  u[k-1], \dots, u[k-(N-1)]\right.)$.
The characteristic function of its $K$th-order joint probability density function $p_{u}\big(u[k], u[k-1], \dots, u[k-(K-1)]\big)$ is given by
\begin{eqnarray}
\label{ref:eqcharacspline}
\hat p_{u}(\omega_1,\dots,\omega_{K})=\Form_w \left(\sum_{k=1}^{K} \omega_k\beta^\vee_{\Lop}(t-k+1) \right)
\end{eqnarray}
where $\beta_{\Lop}$ is the generalized B-spline defined by (\ref{eq:Lspline}). The autocorrelation sequence of the process is compactly-supported:
$$
r_d[k]=\E\left\{u[ k'] \cdot \overline{u[ k'+k]}\right\}=\sigma_0^2 \beta_{\overline{\Lop} \Lop^\ast}(k)
$$
where $\beta_{\overline{\Lop} \Lop^\ast}(t)=(\overline{\beta_{\Lop}} \ast \beta_{\Lop}^\vee)(t)$, while its power spectrum is simply
\begin{eqnarray*}
\Phi_{u}(e^{j\omega})&=&\sigma_0^2 B_{\Lop}(e^{j \omega})
\end{eqnarray*}
where $B_{\Lop}(e^{j \omega})$ is defined by (\ref{eq:DiscreteBsplinekern}).
\end{property}

\begin{proof} The result is a consequence of Proposition \ref{Prop:increment}. 
The pointwise specification (characteristic function of order $K$) is obtained by making the substitution $\varphi=\omega_1\delta(\cdot)+ \cdots + \omega_K \delta(\cdot-K+1)$ in the characteristic form $\Form_{u}(\varphi)=\Form_{w}(\beta_{\Lop}^\vee \ast \varphi)$. The independence between $u[k]$ and $u[k']$ for any $k'$ such that $|k-k'|\ge N$ then follows from the fact that the corresponding B-splines are non-overlapping (since the support of $\beta_{\Lop}$ is of size $N$). 
Indeed, the generic L\'evy noise functional \eqref{eq:levyfunctional} with $f(0)=0$
has the property that $\Form_w(\varphi_1+\varphi_2)=\Form_w(\varphi_1)\cdot \Form_w(\varphi_2)$ whenever $\varphi_1$ and $\varphi_1$ have non-overlapping support, which is synonymous with independence. As for the autocorrelation sequence, it is simply the sampled version of the one given in Property \ref{Prop:correlation_distance}. Likewise, the power spectrum, whose generic form is
\begin{eqnarray*}
\Phi_{u}(e^{j\omega})&=&\sum_{n \inZ} \Phi_{u}(\omega+2 \pi n)\\
&=&\sigma_0^2  \sum_{n \inZ} |\hat \beta_\Lop(\omega+2 \pi n)|^2, 
\end{eqnarray*}
reduces to the finite sum $\sigma_0^2 \sum_{k=-N}^{N}  \beta_{\overline{\Lop} \Lop^\ast}(k)e^{-j \omega k}$, thanks to the compact support of $\beta_{\overline{\Lop} \Lop^\ast}(t)$.
\end{proof}

We would like to emphasize that the statistical characterization of the discrete increment process in Property \ref{Prop:discreteincrement} is complete and that it covers the full class of Gaussian and non-Gaussian stochastic processes specified by the generic stochastic differential equation (\ref{eq:ODE}), including the unstable scenarios which are outside the classical theory of stationary processes. Noteworthy is the omni-presence of the exponential B-spline kernel $\beta_{\Lop}$, which has a fundamental role in all aspects of the characterization. For instance, we observe that the argument $\varphi(t)=\sum_{k=1}^{K} \omega_n\beta^\vee_{\Lop}(t-k+1)$ in the noise functional $\Form_w$ in (\ref{ref:eqcharacspline}) actually corresponds to the generic form of a cardinal exponential spline with the Fourier variables taking over the role of the B-spline coefficients. Likewise, the correlation structure is entirely specified by the integer samples of $\beta_{\overline{\Lop} \Lop^\ast}(t)$ (the autocorrelation of $\beta_{\Lop}$), while the power spectrum is proportional to $B_{\Lop}(e^{j \omega})$, the so-called discrete B-spline filter, which also enters the definition of the spline interpolator in Property \ref{Prop: interpolation}.

The link of course is not coincidental. In spline theory, the construction of B-splines is motivated by the desire to find the shortest possible basis functions to represent a certain family of spline functions. Here, the introduction of the generalized increment process is aimed at producing a derived signal with the simplest possible statistical structure; in particular, the shortest dependency distance.  The proposed solution is optimal in the sense that it achieves the shortest possible order of dependency, as a consequence of the minimal support property of the B-spline. The localization sequence $d_{\boldsymbol \alpha}[k]$ is obviously not arbitrary; the guiding principle is that $\Delta_{\boldsymbol \alpha}$ must have the same null space as $\Lop$ such as to annihilate all the long-ranging exponential/polynomial modes of $\Lop^{-1}$. Concretely, this is achieved by mapping the continuous-domain poles of the system into the discrete-domain zeros of $D_{\boldsymbol \alpha}(z)$ via the exponential map $z=e^s$ (cf. Eq. (\ref{eq:Dlocal})); this also implies that the minimal length of $d_{\boldsymbol \alpha}[k]$ is $N+1$, which puts a lower bound of $N$ on the size of the B-spline. 

\subsection{Discrete innovation models}
Given the fact that the discrete processes $s[k]$ and $u[k]$ are linked through the difference equation (\ref{eq:discrete_Increment}), it is tempting to investigate whether or not it is possible to go one step further and to specify
$s[k]$ through a discrete ARMA-type model. Ideally, we would like to come up with an equivalent discrete-domain innovation model that is easier to exploit numerically than the defining stochastic differential equation (\ref{eq:ODE}).
To that end, we perform the spectral factorization of the discrete B-spline kernel
\begin{eqnarray}
\label{eq:spectfact}
B_\Lop(z)=\sum_{k=-N}^N \beta_{\overline{\Lop} \Lop^\ast}(k)z^{-k}=B^+_\Lop(z)B^-_\Lop(z)
\end{eqnarray}
where $B^+_\Lop(z)=\sum_{k=0}^{N-1}b_\Lop^+[k] z^{-k}=B^-_\Lop(z^{-1})$ specifies a causal finite impulse response (FIR) filter of size $N$. The crucial point for the argument below is that $B^+_\Lop(e^{j \omega})$ (or, equivalently $B_\Lop(e^{j \omega})$ as in Property \ref{Prop: interpolation}) is non-vanishing, which is equivalent to the requirement that $\beta_\Lop$ generates a valid Riesz basis \cite{Unser2005d}. 

\begin{property}[Stochastic difference equation]
\label{Prop:discrete_innovation}
The sampled process of order $N$ with parameters $(\Lop, {\boldsymbol \alpha})$ satisfies the discrete ARMA-type whitening equation
$$
\sum_{n=0}^{N} d_{\boldsymbol \alpha}[k] s[k-n] = \sum_{m=0}^{N-1} b_\Lop^+[k] e[k-m]
$$
where $d_{\boldsymbol \alpha}$ and  $b_\Lop^+$ are defined by (\ref{eq:Dlocal}) and (\ref{eq:spectfact}), respectively. The driving term $e[k]$ is a discrete stationary white noise (white meaning fully decorrelated or with a flat power spectrum). However, $e[k]$ is a valid innovation sequence with {\em independent}, identically-distributed samples
only if the corresponding continuous-domain process is Gaussian, or, in full generality (i.e., non-Gaussian case), if it is a first-order Markov or L\'evy-type process with $N=1$.
\end{property}
\begin{proof}
Since $|B^+_\Lop(e^{j\omega})|=\sqrt{B_\Lop(e^{j\omega})}$ is non-vanishing and a trigonometric polynomial of $e^{j\omega}$ whose roots are inside the unit circle, we have the guarantee that the inverse filter whose frequency response is $\frac{1}{B^+_\Lop(e^{j \omega})}$ is causal-stable. It follows that
$\Phi_{e}(e^{j\omega})=\sigma_0^2  \frac{\sum_{n \inZ}|\hat \beta_\Lop(\omega+2 \pi n )|^2}{B_\Lop(e^{j\omega})} =\sigma_0^2$,
which proves the first part of the statement. As for the second part, we recall that decorrelation is equivalent to independence in the Gaussian case only. In the non-Gaussian case, the only way to ensure independence is by restricting ourselves to a first-order process, which results into an AR(1)-type equation with $e[n]=u[n]$. 
Indeed, Property \ref{Prop:discreteincrement} implies that, for $N=1$,
$p_{u}\left(u[k]\left| \{u[k-m]\}_{m \inZ^+}\right)\right.=p_{u}(u[k])$. This is equivalent to $s[k]$ having the Markov property since
$p_{s}\left(s[k]\left| \{s[k-m]\}_{m \inZ^+}\right)\right.$  $=p_{u}(u[k])$ $=p_{s}\left(s[k]\left| s[k-1]\right)\right.$.
\end{proof}

The fact that continuous-time and discrete-time ARMA models are linked to each other is a classical result in the theory of Gaussian stationary processes \cite{Doob1962}. The present contribution to the topic is: 1) to make the connection completely explicit thanks to the introduction of the localization filter $D_{\boldsymbol \alpha}(z)$ and the discrete B-spline kernel $B_\Lop(z)$, and 2) the extension of the result for the non-stationary and/or non-Gaussian scenarios.
\section{Numerical generation of stochastic processes}
\subsection{Determination of B-splines}
The generalized exponential B-splines were introduced in \cite{Unser2005d} in order to establish a formal link between the continuous-time and discrete-time theories of linear systems. These functions are slightly more general than the classical ones specified by (\ref{eq:BsplineN}), which are missing ``zeros". Since a differential LSI system is characterized by its poles $\boldsymbol \alpha = (\alpha_1,\dots,\alpha_N)$ and zeros $\boldsymbol \gamma = (\gamma_1,\dots,\gamma_M)$ with $M < N$, the idea is to associate it with an identifying exponential B-spline function:
\begin{align}
\beta_{({\boldsymbol \alpha};{\boldsymbol \gamma})}(t)\onetwocol{Ê}{ \hspace{6.4cm}\nonumber\\ }=\Fourier^{-1}\left\{ \left(\prod_{m=1}^M (j \omega-\gamma_m)\right) \prod_{n=1}^N\frac{1-e^{\alpha_n-j\omega}} {j \omega-\alpha_n}\right\}(t).
\label{eq:BsplineE}
\end{align}
Such B-splines can be computed explicitly on a case-by-case basis using the mathematical software described in \cite[Appendix A]{Unser2005d}; Matlab code is also available from the authors on request. The connection with Eq. (\ref{eq:Lspline}) is $\beta_{\Lop}(t)=b_M  \,\beta_{({\boldsymbol \alpha};{\boldsymbol \gamma})}(t)$  where the $\alpha_n$ and $\gamma_m$
are the roots to the polynomial $P_N(\s)= \s^N+ a_{N-1}\s^{N-1}+\dots+a_1 \s + a_0$ and $Q_M(\s)=b_M \s^M+ b_{M-1}\s^{N-1}+\dots+b_1 \s + b_0$, respectively.
The basic operations of the corresponding B-spline calculus are: 
\begin{itemize}
\item Convolution by concatenation of parameter vectors: $(\beta_{({\boldsymbol \alpha}_1;{\boldsymbol \gamma}_1)} \ast \beta_{({\boldsymbol \alpha}_2;{\boldsymbol \gamma}_2)})(t)=\beta_{({\boldsymbol \alpha}_1:{\boldsymbol \alpha}_2;\,{\boldsymbol \gamma}_1:{\boldsymbol \gamma} _2)} (t)$
\item Mirroring by sign change: $\beta_{({\boldsymbol \alpha};{\boldsymbol \gamma})}(-t)=(-1)^M \left( \prod_{n=1}^N e^{\alpha_n} \right)\beta_{(-{\boldsymbol \alpha};-{\boldsymbol \gamma})}(t+N)$
\item Complex-conjugation: $\overline{\beta_{({\boldsymbol \alpha};{\boldsymbol \gamma})} (t)}=\beta_{(\overline{\boldsymbol \alpha};\overline{\boldsymbol \gamma})}(t)$
\item Modulation by parameter shifting: $e^{j \omega_0 t} \beta_{({\boldsymbol \alpha};{\boldsymbol \gamma})} (t)= \beta_{({\boldsymbol \alpha}+{\boldsymbol j}\omega_0;{\boldsymbol \gamma}+{\boldsymbol j}\omega_0)} (t)$
with the convention that ${\boldsymbol j}=(j,\dots,j)$.
\end{itemize}
It follows that the autocorrelation B-spline $\beta_{\overline{\Lop} \Lop^\ast}=\overline{\beta_{\Lop}} \ast \beta_{\Lop}^\vee$ that is central to our formulation is given by
\begin{align}
\beta_{\overline{\Lop} \Lop^\ast}(t)&=b_M^2 \left(\overline{\beta_{({\boldsymbol \alpha};{\boldsymbol \gamma})}} \ast \beta_{({\boldsymbol \alpha};{\boldsymbol \gamma})}^\vee\right)(t) \nonumber \\
&=
b_M^2 (-1)^M \left( \prod_{n=1}^N e^{\alpha_n} \right)
\beta_{(\overline{\boldsymbol \alpha}:-{\boldsymbol \alpha};\,
\overline{\boldsymbol \gamma}:-{\boldsymbol \gamma})}(t+N)
\end{align}

\subsection{Discrete inverse operators}
We have seen that the discrete increment process has a much simpler statistical structure than the process from which it is derived.
This is not only advantageous for the analysis of such stochastic processes, but also exploitable for synthesis purposes. The latter calls for a discrete operator mechanism for inverting the difference equation (\ref{eq:discrete_Increment}). The technique that we propose is in all points analogous to the continuous-domain method presented in Section \ref{sec:solutionODE}.
The principle is to factorize $\Delta_{\boldsymbol \alpha}=\Delta_{\alpha_N} \cdots \Delta_{\alpha_1}$ where each individual operator $\Delta_{\alpha_n}$ actually corresponds to a discrete FIR filter with transfer function $D_{\alpha_n}(z)=1-e^{\alpha_n}z^{-1}$.

Formally, the inverse operator of  $\Delta_{\alpha}$ is the digital filter whose impulse response $h_\alpha[k]$ is the inverse $z$-transform of $\frac{1}{1-e^{\alpha}z^{-1}}=\frac{-e^{-\alpha} z}{1-e^{-\alpha}z}$.
Classical system theory tells us that such a first-order filter is causal-stable iff. its $z$-domain pole 
$z_p=e^{\alpha}$ is inside the unit circle, which is equivalent to ${\rm Re}(\alpha)<0$. It is also possible to change the domain of stability to ${\rm Re}(\alpha)>0$ by switching to an anti-causal response instead of a causal one. The corresponding definition of the impulse response is
\begin{eqnarray*}
h_{\alpha}[k]=\left\{
\begin{array}{ll}
 \One_+[k] e^{\alpha k} & \mbox{if  } {\rm Re}(\alpha)\le 0 \\
- \One_+[-k-1] e^{\alpha k} & \mbox{else} 
\end{array} \right.
\end{eqnarray*}
which is the sampled version of $\rho_\alpha(t)$ in \eqref{eq:rhoalpha} (if one excludes the point of discontinuity of $ \One_+(t)$ at $t=0$). The critical configuration is ${\rm Re}(\alpha)=0$ in which case $h_\alpha[k]$ is still bounded---but not in $\ell_1$---meaning that the filter is no longer stable. 

At any rate, the main point is that $\Delta_{\alpha_n}^{-1}x[k]=(h_{\alpha_n} \ast x)[k]$, and that these first-order inverse filters can be implemented recursively as:

Causal recursion for ${\rm Re}(\alpha_n)\le0$
$$
y[k]=(h_{\alpha_n} \ast x)[k] = e^{\alpha_n}y[k-1] + x[k]
$$

Anti-causal recursion for ${\rm Re}(\alpha_n)>0$
$$
y[k]=(h_{\alpha_n} \ast x)[k] = e^{-\alpha_n}(y[k+1] - x[k+1])
$$
The final ingredient is the discrete counterpart of the operator $\Iop_{\omega_0,\delta}$ specified by (\ref{eq:inv1}); that is, the unique right inverse of $\Delta_{j \omega_0}$ that sets the output signal to zero at $k=0$. This operator, which is denoted by $\Delta_{j \omega_0,\delta}^{-1}$, is given by
$$\Delta_{j \omega_0,\delta}^{-1}\{x\}[k] = (h_{j\omega_0} \ast x)[k] - e^{j \omega_0 k}(h_{j\omega_0} \ast x)[0]$$
where the second term is a properly-weighted complex sinusoid that is in the null space of $\Delta_{j\omega_0}$.
For $k\ge k_0$, the above formula simplifies to
$$
\Delta_{j \omega_0,\delta}^{-1}\{x\}[k] = \sum_{m=0}^k x[m]Êe^{j \omega_0 (k-m)},
$$
which is an expression that can also be updated recursively. 
If $k<0$, the summation bounds are simply interchanged.
Using the same notation and pole ordering as in Section \ref{sec:solutionODE}, we are then able to specify a global right inverse of $\Delta_{\boldsymbol \alpha}$ as
\begin{eqnarray}
\label{eq:invlocdiscrete}
\Delta_{\boldsymbol \alpha}^{-1}=\underbrace{\Delta^{-1}_{j \omega_{n_0}, \delta} \cdots \Delta^{-1}_{j \omega_{1}, \delta} }_{\mbox{shift-variant}}\; \underbrace{\Delta_{\alpha_{N-n_0}}^{-1} \cdots \Delta_{\alpha_1}^{-1} }_{\mbox{ LSI part}},
\end{eqnarray}
which are used to specify the corresponding continuous-domain boundary conditions (\ref{eq:boundary}).
Let $s[k]=\Delta_{\boldsymbol \alpha}^{-1}r[k]$ where $r[k]$ be an arbitrary input signal. Then, the above operator imposes the $n_0$ boundary conditions
\begin{eqnarray}
\label{eq:boundarydiscrete}
\left\{
\begin{array}{rcl}
s[0]&=&0  \\
\Delta_{j \omega_{n_0}}\{s\}[0]&=&0\\
&\vdots&\\
\Delta_{j \omega_2} \cdots \Delta_{j \omega_{n_0}}\{s\}[0]&=&0,
\end{array}
\right.
\end{eqnarray}
while its right-inverse property ensures that $\Delta_{\boldsymbol \alpha}s[k]=\Delta_{\boldsymbol \alpha}\Delta_{\boldsymbol \alpha}^{-1}r[k]=r[k]$. In the stationary case where $n_0=0$ (i.e., ${\rm Re}(\alpha_n)\ne0, n=1,\dots,N$), we also have that $\Delta^{-1}_{\boldsymbol \alpha}\Delta_{\boldsymbol \alpha}r[k]=r[k]$ (left-inverse property).

A small word of caution is in order here. The above discrete-domain boundary conditions are only equivalent to the continuous-domain ones in (\ref{eq:boundary}) for $n_0\le1$. Indeed, it is illusory to attempt imposing exact constraints on the derivatives of such signals if all we have at our disposal are samples on a discrete grid.
The good news, however, is that $\Delta_{j \omega_{n}}$ is, by construction, the best first-order approximation of the continuous-domain operator $\Dop-j \omega_n \Iop$ with the property that : $\Delta_{j \omega_{n}}s(t)=\left(\beta_{j \omega_n} \ast (\Dop-j \omega_n \Identity)s\right)(t)$ where $\beta_{j \omega_n}$ is the corresponding first-order B-spline. In particular, the latter equation ensures convergence to the exact derivatives as the reconstruction grid gets finer (in the same way as finite differences tend to derivatives as the step size goes to zero).

The theoretical alternative is to accept the discrete-domain boundary conditions as they are, assuming that we can properly map them back into
the continuous domain. This is indeed feasible by extending our notion of continuous-domain boundary conditions, as shown in the appendix. The main point is that there is a unique right inverse of $\Lop$ that is admissible (in the sense of  \cite[Theorem 3]{Unser2012}) and compatible with the ``discrete" boundary conditions (\ref{eq:boundarydiscrete}): it is described in the last paragraph of the appendix.

\subsection{Algorithms}
\label{eq:algorithm}
\subsubsection{Gaussian case} The generation of the samples of a generalized Gaussian random process is straightforward since we can rely on the equivalent
discrete innovation (ARMA) model in Property \ref{Prop:discrete_innovation}.
Given a set of parameters ${\boldsymbol \alpha}$ (poles), ${\boldsymbol \gamma}$ (zeros), and $\sigma_0^2$ (noise variance), the procedure is then as follows:
\begin{itemize}
\item Computation of $B_{\Lop}(z)$ and spectral factorization as in (\ref{eq:spectfact}).
\item Generation of the innovation signal $e[k]$ which is a random sequence of i.i.d. Gaussian random variables with zero mean and variance $\sigma_0^2$.
\item FIR filtering with $b_\Lop^+$ and inversion of the model via the application of the inverse operator $\Delta_{\boldsymbol \alpha}^{-1}$ which may be time-invariant or not, depending on the type of process.
\end{itemize}
\subsubsection{Poisson case} This case is slightly more difficult, but can still be handled exactly by starting from the generalized increment process $u[k]$.
Here, we are using the fact that a realization of a Poisson noise with parameter $(\lambda; p_A(a))$ has the explicit form
 $$w(t)=\sum_n a_n\delta(t-t_n)$$
 where $t_n$ are random, uniformly-distributed locations over the real line (point process) with an average density of $\lambda$, and where the amplitudes $a_n$ are i.i.d. random variables with PDF $p_A(a)$.
If we now restrict the observation of the process over a time interval $[0,T]$, the generation may proceed as follows:
\begin{itemize}
\item Analytical computation of the B-spline $\beta_{\Lop}(t)=b_M  \,\beta_{({\boldsymbol \alpha};{\boldsymbol \gamma})}(t)$ using formula (\ref{eq:BsplineE}).
\item Generation of the point process ($t_n$) over the slightly enlarged interval $[-N,T+N]$ together with the amplitude variables $a_n$. This is controlled by first drawing a Poisson-distributed random variable which provides the number of Dirac impulses within the interval.
\item Exact computation of the corresponding discrete increment process by appropriate resampling of the B-spline functions:
$$
u[k]=\left.(\beta_\Lop \ast w)(t)\right|_{t=k}=\sum_{n} a_n \beta_\Lop(k-t_n)
$$
\item Inversion of the model via the application of the inverse operator $\Delta_{\boldsymbol \alpha}^{-1}$ which, again, may be time-invariant or not.
\end{itemize}
In effect, the continuous-time realization of the stochastic process $s$ is a non-uniform $\Lop$-spline with knots at the $t_n$. Its explicit analytical form is
$s(t)=p_0(t)+\sum_{n} a_n \rho_\Lop(t-t_n)$ where $p_0(t)$ is a component that is in the null space of $\Lop$ and $\rho_\Lop(t)$ is a Green function of $\Lop$. In the stationary scenario, $p_0(t)$ may be seen as a random component that condenses all impulsive noise contributions from outside the generation interval. In the non-stationary
case, it has the stricter role of enforcing the $n_0$ boundary conditions imposed by the presence of poles on the imaginary axis.

\subsubsection{Alpha-stable case} Here, we can benefit from the key property that any filtered version of an alpha-stable innovation remains alpha-stable. Indeed, the characteristic function of the variable $X=\langle w, \varphi \rangle$ where $w$ is an S$\alpha$S noise (cf.  specification of $f_\alpha(\omega)$ in Section \ref{sec:charfunc}) is given by
\begin{eqnarray*}
\E\{e^{- j \omega \langle w, \varphi \rangle}\}&=&\Form_w(\omega \varphi)\\
&=&\exp\left(- b_\alpha \|\omega \varphi \|_{L_\alpha}^\alpha \right) \\
&=&\exp\left(- b_\alpha \|\varphi\|_{L_\alpha}^\alpha \cdot |\omega|^\alpha \right)
\end{eqnarray*}
where $\|\varphi\|_{L_\alpha}^\alpha=\int_\R |\varphi(t)|^\alpha \dint t$ is a normalization constant that is shift-invariant; that is, $\|\varphi(\cdot - t_0)\|_{L_\alpha}^\alpha=\|\varphi\|_{L_\alpha}^\alpha$. This implies that $X$ has an alpha-stable distribution, and by extension, that any linear transformation of an alpha-stable process $s$ is alpha-stable as well \cite{Feller1971,Samorodnitsky1994}. It is therefore a simple matter to generate an alpha-stable Markov process ($N=1$) whose increments are independent (cf. Property \ref{Prop:discrete_innovation}).
The situation gets more delicate for higher-order processes because of the necessity of generating an alpha-stable discrete increment sequence with an $N$th-order of dependency. 
The first approach that comes to mind is to run an adapted version of the Gaussian algorithm where the discrete input innovation is alpha-stable instead of Gaussian. Since alpha-stable laws are preserved through linear combinations, this will at least ensure that the marginals are alpha-stable and that the second-order dependencies are the correct ones. This discrete innovation approach, however, is not entirely satisfactory because decorrelation is not rigorously equivalent to independence.

The alternative approach that we propose is to use a piecewise-constant approximation of the B-spline $\beta_\Lop$ with an oversampling factor of $m$:
$$
\beta_{\Lop, m}(t)=\sum_{k=0}^{mN} \beta_\Lop(k/m)\beta_0\left(m(t - \frac{k}{m})\right).
$$
where $\beta_0(mt)=1_{[0,\frac{1}{m})}(t)$ is a rectangular function of size $1/m$.
The basic results from approximation theory ensure that $\lim_{m \rightarrow \infty} \beta_{\Lop, m}(t)= \beta_\Lop(t)$ pointwise and in all $L_p$-norms with the error decaying like $1/m$ (since piecewise-constant splines have first-order of approximation). Starting from the oversampled version of the first-order alpha-stable increment process $s_{d,1}(k/m)=\big(\beta_0(m\cdot) \ast w\big)(k/m)$, which is an i.i.d. alpha-stable sequence, we are then able to compute the samples of the discrete increment process by applying the following convolution-like equation
$$
\left.(\beta_{\Lop, m} \ast w)(t)\right|_{t=k'}=\sum_{k=0}^{mN} \beta_\Lop(k/m)s_{d,1}\left(m(k' - \frac{k}{m})\right).
$$ 
The approximation can be made arbitrary close by increasing the over-sampling factor $m$. The computational overhead is essentially that of generating $m$ times more i.i.d. random variables as in the Gaussian algorithm. The remainder of the procedure is the same as in the Poisson case. Note that this algorithm is generic and applicable to other types of L\'evy innovations as well. 

We conclude this section by indicating that we can also arbitrarily change the sampling step (which had been set to $T=1$ for simplicity) via a simple rescaling of the poles, zeros and noise variance. The main point of the argument is that $\hat L_{({\boldsymbol \alpha};{\boldsymbol \gamma})}(T \omega)=T^{M-N} \hat L_{({\boldsymbol \alpha}/T;{\boldsymbol \gamma}/T)}(\omega)$ and that the white noise property is invariant to dilation (up to a normalization factor).
\section{Illustrative Examples}

\begin{figure}
\centering
\onetwocol{ 
 \includegraphics[scale=0.66]{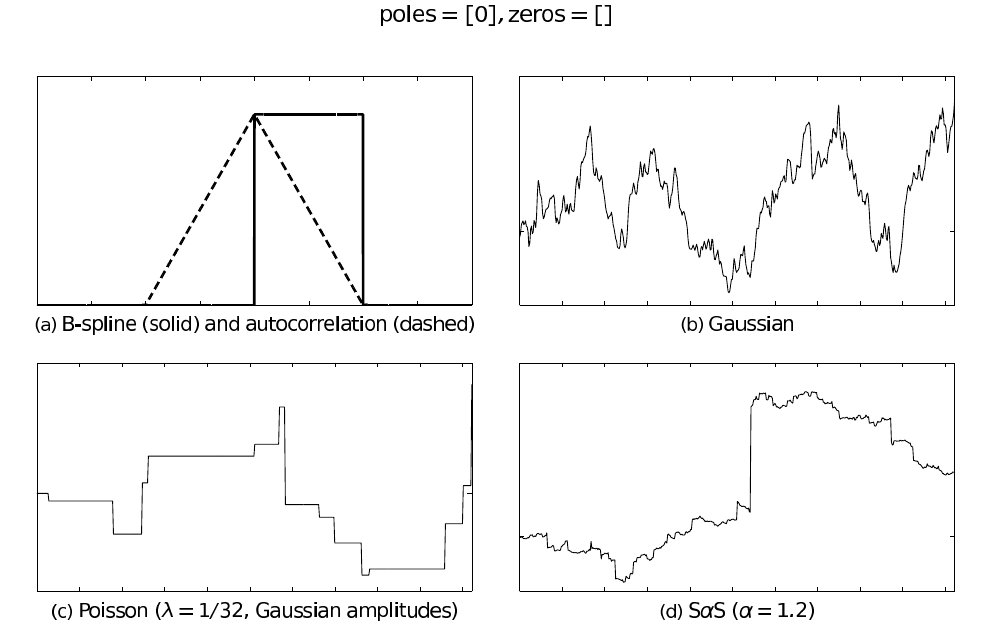}
 }{
 \includegraphics[scale=0.53]{S2Fig2.pdf}
\vspace*{-3ex} }
  \caption{\label{fig:example1}
Example 1: Generation of generalized stochastic processes with whitening operator $\Lop=\Dop$ \big(pole vector $\boldsymbol \alpha=(0)$\big): (a) B-spline functions $\beta_\Lop(t)={\rm rect}\big(t-\frac{1}{2}\big)$ and $\beta_{\overline{\Lop}\Lop^\ast}(t)={\rm tri}(t)$, (b) Brownian motion, (c) Compound Poisson process with $\lambda=1/32$ and Gaussian amplitude distribution $p_A(a)=(2 \pi)^{-1/2} e^{-a^2/2}$, c) S$\alpha$S L\'evy motion with $\alpha=1.2$.}
\end{figure}

\begin{figure}
\centering
\onetwocol{ 
 \includegraphics[scale=0.66]{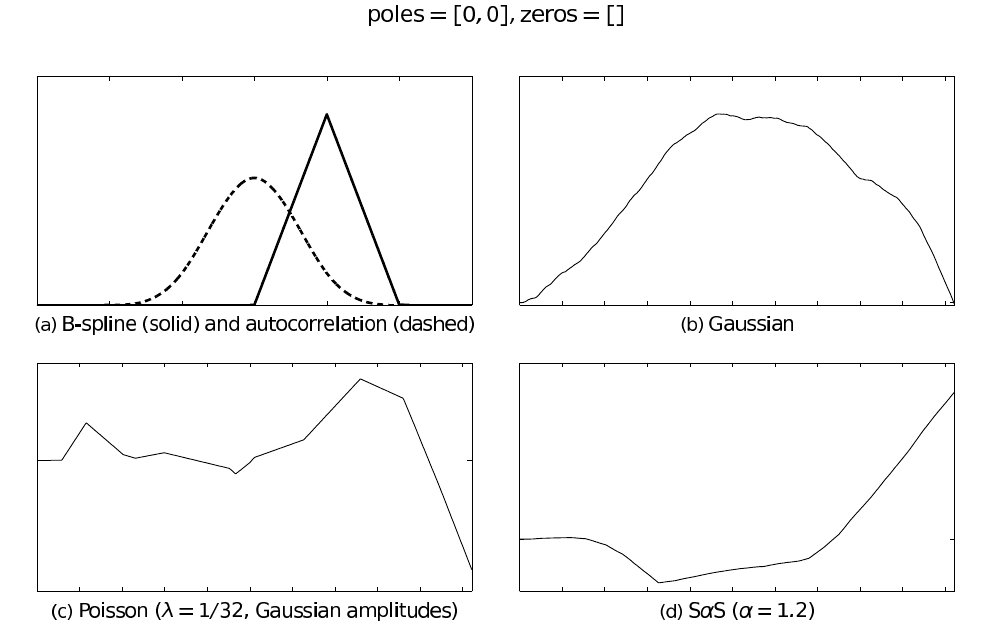}
 }{
 \includegraphics[scale=0.53]{S2Fig3.pdf}
\vspace*{-3ex} }
  \caption{\label{fig:example2}
Example 2: Generation of generalized stochastic processes with whitening operator $\Lop=\Dop^2$ \big(pole vector $\boldsymbol \alpha=(0,0)$\big): (a) B-spline functions $\beta_\Lop(t)={\rm tri}(t)$ and $\beta_{\overline{\Lop}\Lop^\ast}(t)$  (cubic B-spline), (b) Gaussian process, (c) generalized Poisson process with $\lambda=1/32$ and Gaussian amplitude distribution, c) generalized S$\alpha$S process with $\alpha=1.2$.}
\end{figure}

Examples of realizations of Gaussian versus sparse stochastic processes are shown in Figs. 2 to 5. 
These signals were generated using the algorithms described in Section \ref{eq:algorithm} for the three types of driving noises: Gaussian (panel b), impulsive Poisson (panel c), and symmetric-alpha-stable (S$\alpha$S) with $\alpha=1.2$ (panel d).

The relevant operators are: 
\begin{itemize}
\item Example 1: $\Lop=\Dop$ (L\'evy process)
\item Example 2: $\Lop=\Dop^2$  (second-order extension of L\'evy process)
\item Example 3: $\Lop=(\Dop-\alpha_1\Identity)(\Dop-\alpha_2\Identity)$ and $\boldsymbol \alpha=(j 3\pi/4,-j 3\pi/4)$ (generalized L\'evy process)
\item Example 4: $\Lop=(\Dop-\alpha_1\Identity)(\Dop-\alpha_2\Identity)$ and $\boldsymbol \alpha=(-0.05+j \pi/2,-0.05-j \pi/2)$ (CAR(2) process)
\end{itemize}
The corresponding B-splines ($\beta_\Lop$ and $\beta_{\overline{\Lop}\Lop^\ast}$) are shown in the upper left panel of each figure. 

The signals that are displayed side-by-side share the same whitening operator, but they differ in their sparsity patterns which come in three flavors: none (Gaussian), finite rate of innovation (Poisson), and heavy-tailed statistics (S$\alpha$S).
The Gaussian signals are uniformly textured, while the generalized Poisson ones are piecewise-smooth by construction. 

\subsection{Self-similar processes}
The classical L\'evy processes (Fig. 2) are obtained by integration of white L\'evy innovation; they go hand-in-hand with the B-spline of degree 0 (rect), and its autocorrelation (triangle function) which is a B-spline de degree 1. The Gaussian version (Fig. 2b) is a Brownian motion. It is quite rough and nowhere differentiable in the classical sense. Yet, it is mean-square continuous due to the presence of the single pole at the origin. The Poisson version (compound Poisson process) is piecewise-constant, each jump corresponding to the occurrence of a Dirac impulse. The S$\alpha$S L\'evy motion exhibits local fluctuations punctuated by large (but rare) jumps, as is characteristic for this type of process\cite{Samorodnitsky1994,Appelbaum2009}. Overall, it is the jump behavior that dominates making it even sparser than its Poisson counterpart.

The  example in Fig. 3 (second-order extension of a L\'evy process) corresponds to one more level of integration which yields smoother signals (i.e., one-time differentiable in the classical sense). The corresponding Poisson process is piecewise-linear, while the S$\alpha$S version looks globally smoother than the Gaussian one, except for a few sharp  discontinuities in its slope. The basic B-spline here is a triangle, while $\beta_{\overline{\Lop}\Lop^\ast}$ is a cubic B-spline.
The signals in Fig. 2 and 3 are non-stationary;  the underlying processes have the remarkable property of being self-similar (fractals) due to the scale-invariance of the pure derivative operators. The Gaussian and S$\alpha$S stable processes are strictly self-similar in the sense that the statistics are preserved through rescaling. By contrast, the scaling of the Poisson processes necessitates some corresponding adjustment of the rate parameter $\lambda$ \cite{Unser2011}.

\begin{figure}
\centering
\onetwocol{ 
 \includegraphics[scale=0.66]{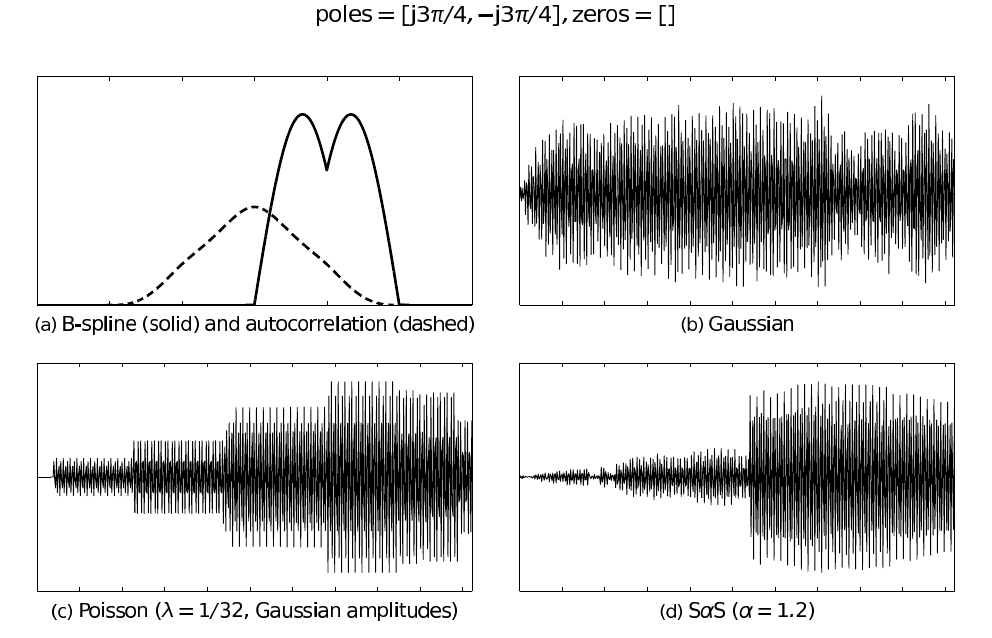}
 }{
 \includegraphics[scale=0.53]{S2Fig4.pdf}
\vspace*{-3ex} }
  \caption{\label{fig:example3}
Example 3: Generation of generalized stochastic processes with whitening operator $\Lop=(\Dop-\alpha_1\Identity)(\Dop-\alpha_2\Identity)$ and $\boldsymbol \alpha=(j 3\pi/4,-j 3\pi/4)$: (a) B-spline functions $\beta_\Lop$ and $\beta_{\overline{\Lop}\Lop^\ast}$, (b) Gaussian process, (c) generalized Poisson process with $\lambda=1/32$ and Gaussian amplitude distribution, c) Generalized S$\alpha$S process with $\alpha=1.2$.} 
\end{figure}

\begin{figure}
\centering
\onetwocol{ 
 \includegraphics[scale=0.66]{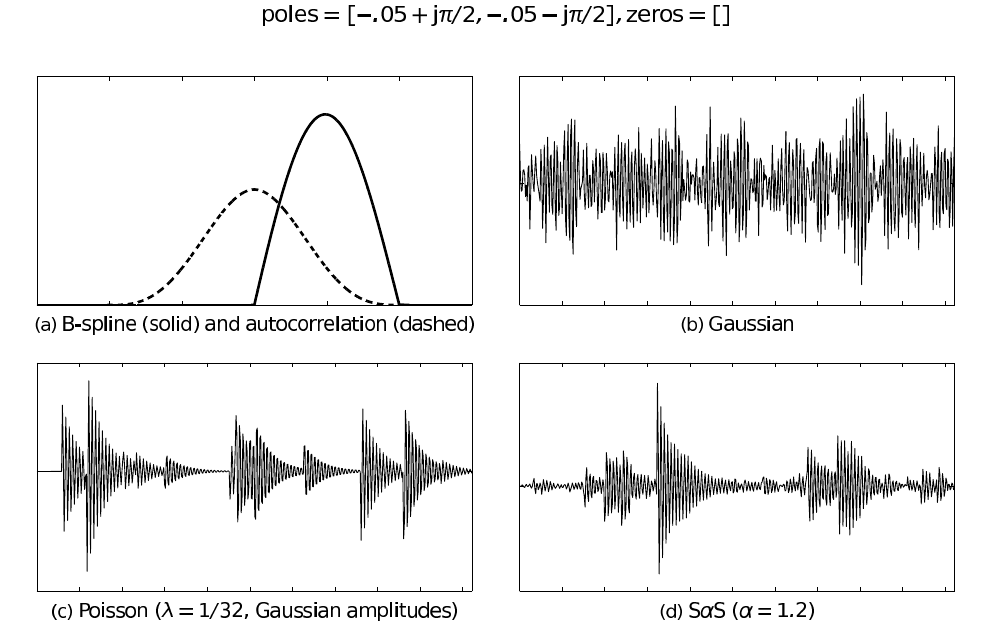}
 }{
 \includegraphics[scale=0.53]{S2Fig5.pdf}
\vspace*{-3ex} }
  \caption{\label{fig:example4}
Example 4: Generation of generalized stochastic processes with whitening operator $\Lop=(\Dop-\alpha_1\Identity)(\Dop-\alpha_2\Identity)$ and $\boldsymbol \alpha=(-0.05+j \pi/2,-0.05-j \pi/2)$: (a) B-spline functions $\beta_\Lop$ and $\beta_{\overline{\Lop}\Lop^\ast}$, (b) Gaussian AR(2) process, (c) Generalized Poisson process with $\lambda=1/32$ and Gaussian amplitude distribution, c) S$\alpha$S AR(2) process with $\alpha=1.2$. }
\end{figure}
\subsection{Bandpass processes}
The second-order signals in Fig. 4 are are non-stationary as well, but no longer self-similar.
They are real-valued, and $C^1$-continuous almost everywhere (pair of complex-conjugate poles in the left complex plane). They constitute some kind of modulated (or bandpass) counterpart of the L\'evy processes which appears to be much better suited for the modeling of acoustic signals. As in the other examples, the Gaussian version is looking cluttered.
The Poisson signal is somewhat stereotyped (stretches of pure oscillating regime) and not quite as realistic looking as its S$\alpha$S counterpart.

As soon as the poles are moved away from the imaginary axis, the processes become stationary.
This is illustrated in Fig. 5 with some CAR(2) (continuous autoregressive) examples, the non-Gaussian versions of which having a marked tendency to exhibit characteristic bursts associated with the impulse response of the system. These latter processes are part of the stationary CARMA family characterized by Brockwell using an alternative stochastic integration/state-space formulation\cite{Brockwell2001}.

\subsection{Mixed processes}
One can also construct signals with a more complex structure by simple addition of independent elementary processes. This results into a mixed process, $s_{\rm mix}=s_1 + \cdots + s_M$, whose characteristic form is the product of the characteristic forms of the individual constituents: 
\begin{eqnarray*}
\Form_{s_{\rm mix}}(\varphi)=\prod_{m=1}^M \Form_{s_m}(\varphi)
=\exp\left( \int_{\R} \sum_{m=1}^M f_m\big(\Lop_m^{-1\ast}\varphi(t)\big) \dint t\right)
\end{eqnarray*}
where $s_m$ is some elementary process with whitening operator $\Lop_m$ and L\'evy exponent $f_m(\omega)$.
As a demonstration of concept, we have synthesized some acoustic samples by mixing random signals associated with elementary musical notes (pair of poles at the corresponding frequency). These can be downloaded from the web at 
{\tt http://bigwww.epfl.ch/sparse}. The Gaussian versions are diffuse, cluttered and boring to listen to. Our generalized Poisson and S$\alpha$S samples are more interesting perceptually---reminiscent of chimes---with the latter sounding less dry and more realistic. Note that mixing does not gain us anything in the Gaussian case because the resulting signal is still part of the traditional family of Gaussian ARMA processes (this follows from Parseval's relation and the fact that $\sum_{m=1}^M \frac{\sigma_0^2}{|\hat L_m(-\omega)|^2}$ is expressible as an equivalent rational power spectrum). This is not so for the non-Gaussian members of the family, which are generally not decomposable, meaning that the mixing of sparse processes opens up new modeling perspectives.
Interestingly, the Gaussian acoustic samples are almost impossible to compress using mp3/AAC, while the generalized Poisson and S$\alpha$S ones can be faithfully reproduced at a much lower bit rate.

\section{Conclusion}

The main point of this paper has been to show that the spline interpretation that links the continuous- and discrete-time deterministic linear system theories has a direct counterpart in the linear theory of stochastic processes. 
While the connection between SDEs and stochastic difference equations is well understood in the classical framework of Gaussian stationary processes, it is much less so when (i) the excitation noise is non-Gaussian, and/or (ii) when the underlying system is unstable. We have argued that these two extensions are essential for producing signals that are sparse---which calls for non-Gaussian excitations---and compressible in a wavelet basis (because self-similar processes are solutions of unstable SDEs). Our main effort in this series of papers has been to address these issues by setting the foundation of a general framework that extends the bounds of the traditional theory of Gaussian stationary processes.
The good news is that our generalized formulation leads to a simple universal conversion scheme by which a stochastic differential equation is mapped into some corresponding stochastic finite difference equation. The cornerstone of this approach is the existence of a compactly supported exponential B-spline, $\beta_\Lop$,
which acts as the mathematical translator between the continuous domain operator $\Lop$ and its discrete version $\Lop_{\rm d}$. 
The elucidation of this A-to-D connection has direct implications for signal synthesis (generation of sparse stochastic processes) and statistical analysis (proper specification of likelihood functions, optimal signal estimation). Most importantly, it provides a functional approach that facilitate the derivation of the joint statistics of such processes, especially in the non-Gaussian cases.

While the proposed framework opens up new modeling perspectives, it also calls for further mathematical investigations.
In particular, more work is required to quantify the sparsifying properties of wavelet-like expansions and to investigate the existence of optimal representations for non-Gaussian processes.
We are also postulating that the smoothness properties (H\"older and Sobolev exponents) of our extended family of CARMA processes are directly related to those of the underlying B-splines. While this is justifyable in the Gaussian and Poisson cases \cite{Adler1981,Unser2011}, the details still need to be worked out for the other brands of innovation, especially the ones with unbounded variance (e.g., S$\alpha$S) for which a mean-square interpretation cannot be provided.
%
\appendices
\section*{Appendix: Generalized boundary conditions}
The guiding principle for defining non-stationary processes with generalized boundary conditions is to extend the class of inverse operators considered in 
\cite[Section III-B]{Unser2012}.
To that end, we introduce the linear operator
\begin{eqnarray}
\label{eq:invgen}
\Iop_{\omega_0,\varphi_0}f(t) &=& \Iop_{\omega_0}f (t) - e^{j \omega_0 t}\;\frac{\langle \Iop_{\omega_0}f, \varphi_0\rangle}{\hat \varphi_0(-\omega_0)},
\end{eqnarray}
where $\Iop_{\omega_0}$ is the traditional shift-invariant inverse operator specified by the inverse Fourier integral 
$$
\Iop_{\omega_0}f(t)=\int_\R \hat f(\omega)\left(\frac{1}{j(\omega-\omega_0)} +\pi \delta(\omega-\omega)\right)e^{j \omega t} \frac{ \dint \omega}{2 \pi},
$$
and where $\varphi_0(t)$ is some given compactly-supported function such that $\hat \varphi_0(-\omega_0)\ne 0$. We note that the above operator is well-defined pointwise for any $f \in L_1$ and that it is a right inverse of $(\Dop-j \omega_0\Identity)$ because the sinusoidal correction on the right 
is in the null space of the operator. By design, $\Iop_{\omega_0,\varphi_0}$ is such that it imposes the generalized boundary condition
\begin{eqnarray}
\label{eq:genboundary}
{\langle \Iop_{\omega_0,\varphi_0}f, \varphi_0\rangle}=0
\end{eqnarray}
for any input function $f$. 

Our next task is to show that the adjoint of this operator is admissible.
To identify $\Iop_{\omega_0,\varphi_0}^\ast$, we perform the inner-product manipulation
\begin{align}
\langle \Iop_{\omega_0,\varphi_0}f, g \rangle&=\langle\Iop_{\omega_0}f , g \rangle - \langle e^{j \omega_0 t}, g \rangle\frac{\langle \Iop_{\omega_0}f, \varphi_0\rangle}{\hat \varphi_0(-\omega_0)}  \nonumber \\
   &=\langle f,\Iop_{\omega_0}^\ast g \rangle - \hat g(-\omega_0) \frac{\langle f, \Iop_{\omega_0}^\ast\varphi_0\rangle}{\hat \varphi_0(-\omega_0)} \nonumber \end{align}
which, by identification with $\langle f, \Iop^\ast_{\omega_0,\varphi_0}g\rangle$, yields
\begin{eqnarray}
\Iop^\ast_{\omega_0,\varphi_0} g(t) &=& \Iop^\ast_{\omega_0}\left\{ g-\frac{\hat g(-\omega_0)}{\hat \varphi_0(-\omega_0)} \varphi_0\right \}(t) 
\label{eq:invgenast}
\end{eqnarray}
where $ \Iop^\ast_{\omega_0}$ is the anti-causal convolution operator whose impulse response is $\rho_{j \omega}^\vee(t)= \One_+(-t)e^{- j \omega_0 t}$. The right-inverse property of $\Iop_{\omega_0,\varphi_0}$ automatically gets transposed into a left-inverse property for its adjoint $\Iop^\ast_{\omega_0,\varphi_0}$.
Next, by using the fact that $\int_{-\infty}^{+\infty}e^{ j \omega_0 t} \left( f(t)-\frac{\hat f(-\omega_0)}{\hat \varphi_0(-\omega_0)} \varphi_0(t) \right) \dint t = 0$
and applying the same technique as in the proof of \cite[Proposition 2]{Unser2012}, we show that
$$
\left|\Iop^\ast_{\omega_0,\varphi_0} f(t)\right| < C_{\varphi_0} \frac{ \|f\|_{\infty,r}}{1 + |t|^{r-1}}
$$
for any $f \in L_{\infty,r}$ (the space of functions with algebraic decay of order $r$) where $C_{\varphi_0}$ is a constant that solely depends upon $\varphi_0$. This proves that $\Iop^\ast_{\omega_0,\varphi_0}$ is a continuous operator on $\mathcal{R}$ (the space of rapidly-decreasing functions), and, by implication, a continuous map from $\mathcal{S}$ into $L_p$ with $p\ge1$. The same holds true for any combination (iteration) of such elementary operators.

For completeness, we are giving the equivalent\footnote{The derivation of the first formula relies on the duality-product version of Parseval's relation: $\langle s, \varphi_0 \rangle=\int_\R s(t) \varphi_0 (t) \dint t=
\frac{1}{2 \pi} \int_\R \hat s(\omega) \hat \varphi_0(-\omega) \dint \omega$ where $\hat \varphi_0(-\omega)=\overline{\Fourier\{\overline{\varphi_0}\}}$.} Fourier-based definition of the relevant pair of inverse operators which are valid for distributions as well:
\begin{align}
\label{eq:invopgen}
\Iop_{\omega_0,\varphi_0}f(t) =\int_{\R}   \hat{f}(\omega) \left(
   \frac{e^{j \omega t}-e^{j \omega_0 t} \frac{\hat \varphi_0(-\omega)}{\hat \varphi_0(-\omega_0)} }
       {j(\omega-\omega_0)} \right)
    \frac{\dint{\omega\;\;}}{2 \pi}\\
    \Iop^\ast_{\omega_0,\varphi_0}f(t) =\int_{\R}    \left(
   \frac{\hat{f}(\omega)-\frac{\hat f(-\omega_0)}{\hat \varphi_0(-\omega_0)} \hat \varphi_0(\omega)}
       {-j(\omega+\omega_0)} \right)e^{j \omega t} 
    \frac{\dint{\omega\;\;}}{2 \pi}.
\label{eq:invop*}
    \end{align}
Observe that both Fourier integrals are non-singular and that we recover the formulas in \cite[Table 1]{Unser2012}, as well as (\ref{eq:inv1}), by setting $\varphi_0=\delta(\cdot-t_0)$ and $\varphi_0=\delta$, respectively.

We can now replicate the construction of an admissible left-inverse operator $\Lop^{-1\ast}$ for the general $N$th-order differential system in 
\cite[Section IV-C]{Unser2012}.
In the case of an $n_0$th-order of singularity, the generic form of a proper inverse operator that is admissible in the sense of (\ref{eq:stable})
 is
\begin{align}
\label{eq:ginvadj}
\Lop^{-1\ast}= \Top^\ast_{\rm LSI}  \Iop^\ast_{\omega_1,\varphi_1}  \cdots \Iop^\ast_{\omega_{n_0},\varphi_{n_0}}
\end{align}
where $\Top_{\rm LSI}$ is some ``standard" $\mathcal{S}$-continuous convolution operator.
The adjoint $\Lop^{-1}=\Iop_{\omega_{n_0},\varphi_{n_0}} \cdots \Iop_{\omega_1,\varphi_1}\Top_{\rm LSI}$, which is the right-inverse of $\Lop$, is then such that it imposes the generalized boundary conditions on the output
signal $s=\Lop^{-1}w$ 
\begin{eqnarray}
\label{eq:boundarygen}
\left\{
\begin{array}{rcl}
\langle \varphi_{n_0}, s\rangle&=&0  \\
\langle \varphi_{n_0-1}, (\Dop-j \omega_{n_0}\Identity)s\rangle&=&0\\
&\vdots&\\
\langle \varphi_{1}, (\Dop-j \omega_{2}\Identity)\cdots(\Dop-j \omega_{n_0}\Identity) s\rangle&=&0,\end{array}
\right.
\end{eqnarray}
for any driving term $w$. 

Interestingly, if we select $\varphi_{n_0}=\delta$, $\varphi_{n_0-1}=\beta^\vee_{j\omega_{n_0}}$, $\varphi_{n_0-2}=\beta^\vee_{(j\omega_{n_0-1}, j\omega_{n_0})}$, \dots, $\varphi_{1}=\beta^\vee_{(j\omega_{2},\cdots  j\omega_{n_0})}$, we end up with a set of continuous-time boundary conditions 
(\ref{eq:boundarygen})
that is rigorously equivalent to the ``discrete" one in (\ref{eq:boundarydiscrete}).
Since the specification of boundary conditions is somewhat arbitrary anyway, this is clearly our preferred choice. It has the advantage of ensuring a perfect compatibility between the continuous and discrete-domain specifications of these processes.


\section*{Acknowledgements}
The research was partially supported by the Swiss National Science Foundation 
under Grant 200020-109415 and by the European Commission under Grant ERC-2010-AdG
267439-FUN-SP. 
\bibliographystyle{IEEEtran}
\bibliography{/Users/munser/Bibliography/Bibtex_files/Unser}

\end{document}